\documentclass[smallextended]{svjour3}       % onecolumn (second format)

\usepackage{graphicx}
\usepackage{amsfonts} % if you want blackboard bold symbols e.g. for real numbers
\usepackage{epsf,  amsmath, amssymb, graphicx}

\usepackage{amsthm}
\usepackage[numbers]{natbib}
\usepackage{dsfont}
\usepackage{bm}
\usepackage{mathtools}
\usepackage{xcolor}
\usepackage{hyperref}
\newcommand{\iid}{\overset{i.i.d.}{\sim}}
\theoremstyle{definition}  % gives non italic without it, Latex gives the default which is \theoremstyle{plain}

\smartqed  % flush right qed marks, e.g. at end of proof
\begin{document}

\title{Large-Scale Kernel Methods for Independence Testing}



\author{ 	Qinyi Zhang \and
		Sarah Filippi \and\\
		Arthur Gretton \and
		Dino Sejdinovic
}


\institute{Qinyi Zhang \at
           Department of Statistics, University of Oxford \\
           \email{qinyi.zhang@stats.ox.ac.uk}          
           \and
           Sarah Filippi \at
           Department of Statistics, University of Oxford \\
           \email{filippi@stats.ox.ac.uk}
           \and
           Arthur Gretton  \at
           Gatsby Computational Neuroscience Unit, University College of London \\
           \email{arthur.gretton@gmail.com}
           \and
           Dino Sejdinovic \at
           Department of Statistics, University of Oxford \\
           \email{dino.sejdinovic@stats.ox.ac.uk}
}

\date{Received: date / Accepted: date}

\maketitle

\begin{abstract}
Representations of probability measures in reproducing kernel Hil-bert spaces provide a flexible framework for fully nonparametric hypothesis tests of independence, which can capture any type of departure from independence, including nonlinear associations and multivariate interactions. However, these approaches come with an at least quadratic computational cost in the number of observations, which can be prohibitive in many applications. Arguably, it is exactly in such large-scale datasets that capturing \emph{any type} of dependence is of interest, so striking a favourable tradeoff between computational efficiency and test performance for kernel independence tests would have a direct impact on their applicability in practice. In this contribution, we provide an extensive study of the use of large-scale kernel approximations in the context of independence testing, contrasting block-based, Nystr\" om and random Fourier feature approaches. Through a variety of synthetic data experiments, it is demonstrated that our novel large scale methods give comparable performance with existing methods whilst using significantly less computation time and memory. 


\end{abstract}

\section{Introduction}

Given a paired sample $\mathbf{z} = \{ (x_i,y_i) \}^m_{i=1}$ with each $(x_i, y_i)\in \mathcal X \times \mathcal Y$ independently and identically following the joint distribution $P_{XY}$ on some generic domains $\mathcal X$ and $\mathcal Y$, the nonparametric independence problem consists of  testing if we should reject the null hypothesis $\mathcal{H}_0: P_{XY} = P_X P_Y$ in favour of the general alternative hypothesis $\mathcal{H}_1: P_{XY} \not = P_X P_Y$, where $P_X$ and $P_Y$ are the marginal distributions for $X$ and $Y$ respectively. This problem is fundamental and extensively studied, with wide-ranging applications in statistical inference and modelling. Classical dependence measures, such as Pearson's product-moment correlation coefficient, Spearman's $\rho$, Kendall's $\tau$ or methods based on contingency tables are typically designed to capture only particular forms of dependence (e.g. linear or monotone). Furthermore, they are applicable only to scalar random variables or require space partitioning limiting their use to relatively low dimensions. As availability of larger datasets also facilitates building more complex models, dependence measures are sought that capture more complex dependence patterns and those that occur between multivariate and possibly high-dimensional datasets. In this light, among the most popular dependence measures recently have been those based on characteristic functions \cite{Cordis2007,SR2009} as well as a broader framework based on kernel methods \cite{gretbousmol2005,greker08}. A desirable property of consistency against any alternative - i.e. test power provably increasing to one with the sample size regardless of the form of dependence, is warranted for statistical tests based on such approaches. However, this is achieved at an expense of computational and memory requirements that increase at least quadratically with the sample size, which is prohibitive for many modern applications. Thus, a natural question is whether a favourable tradeoff between computational efficiency and test power can be sought with appropriate large-scale approximations. As we demonstrate, several large-scale approximations are available in this context and they lead to strong improvements in power-per-computatonal unit performance, resulting in a fast and flexible independence testing framework responsive to all forms of dependence \emph{and} applicable to large datasets.\\

The key quantity we consider is the Hilbert-Schmidt Independence Criterion (HSIC) introduced by Gretton et al \cite{gretbousmol2005}. HSIC uses the distance between the kernel embeddings of probability measures in the Reproducing Kernel Hilbert Space (RKHS) \cite{greker08,ZhPeJanSch11,probembedding}. By building on decades of research into kernel methods for machine learning \cite{scholkopf2002learning}, HSIC can be applied to multivariate observations as well as to those lying in non-Euclidean and structured domains, e.g., \cite{greker08} considers independence testing on text data. HSIC has also been applied to clustering and learning taxonomies \cite{Song2007,Blaschko2009}, feature selection \cite{SonSmoGreBedetal12}, causal inference \cite{Peters2014,Flaxman2015,Zaremba2014} and computational linguistics \cite{Nguyen2016}. A closely related dependence coefficient that measures all types of dependence between two random vectors of arbitrary dimensions is the distance covariance (dCov) of \cite{Cordis2007,SR2009}, which measures distances between empirical characteristic functions or equivalently measures covariances with respect to a stochastic process \cite{SR2009}, and its normalised counterpart, distance correlation (dCor). RKHS-based dependence measures like HSIC are in fact extensions of dCov -- \cite{SejSriGreFuku13} shows that dCov can be understood as a form of HSIC with a particular choice of kernel. Moreover, dCor can be viewed as an instance of kernel matrix alignment of \cite{Cortes2012}. As we will see, statistical tests based on estimation of HSIC and dCov are computationally expensive and require at least $\mathcal{O}(m^2)$ time and storage complexity, where $m$ is the number of observations, just to compute an HSIC estimator which serves as a test statistic. In addition, the complicated form of the asymptotic null distribution of the test statistics necessitates either permutation testing \citep{arcones1992} (further increasing the computational cost) or even more costly direct sampling from the null distribution, requiring eigendecompositions of kernel matrices using the spectral test of \cite{GreFukHarSri09}, with a cost of $\mathcal{O}(m^3)$. These memory and time requirements often make the HSIC-based tests infeasible for practitioners. 
%The resulting hypothesis test is consistent against all alternatives provided that the variables have bounded first moment. 

%  Another closely related concept is the kernel conditional dependence measure proposed by \cite{Fukumizu08kernelmeasures}, which encodes the dependency structure of the variables in the Hilbert Schmidt norm of the normalised conditional cross covariance operator. It includes unconditional dependence as a special case, i.e. the Hilbert Schmidt norm of the normalised cross covariance operator (NOCCO). In fact, it can be seen as HSIC with whitened features. \\

In this paper, we consider several ways to speed up the computation in HSIC-based tests. More specifically, we introduce three novel fast estimators of HSIC: the block-based estimator, the Nystr\"om estimator and the random Fourier feature (RFF) estimator and study the resulting independence tests.  In the block-based setting, we obtain a simpler asymptotic null distribution as a consequence of the Central Limit Theorem in which only asymptotic variance needs to be estimated - we discuss possible approaches for this. RFF and Nystr\"om estimators correspond to the primal finite-dimensional approximations of the kernel functions and as such also warrant estimation of the null distribution in linear time -- we introduce spectral tests based on eigendecompositions of primal covariance matrices, which avoid permutation approach and significantly reduce the computational expense for the direct sampling from the null distribution.

%first, we study a novel RFF spectral approach which estimates the eigenvalues of the integral kernel operator appearing in the null distribution using RFF and hence reduces computation in the spectral approach of \cite{GreFukHarSri09}. Second, we note that a permutation approach as  can also be adopted. For the case of the Nystr\"om estimator, these two approaches can still be applied with the explicit feature map representation given by the Nystr\"om approximation.\\

\subsection*{Related Work}

Some of the approximation methods considered in this paper were inspired by their use in a related context of two-sample testing. In particular, the block-based approach for two-sample testing was studied in \cite{Gretton2sample,optikernel12,ZarGreBla13} under the name of linear-time MMD (Maximum Mean Discrepancy), i.e. the distance between the mean embeddings of the probability distributions in the RKHS. The approach estimates MMD on a small block of data and then averages the estimates over blocks to obtain the final test statistic. Our block-based estimator of HSIC follows exactly the same strategy. On the other hand, 
The Nystr\"om method \cite{Nystrom, Snelson06SG} is a classical low-rank kernel approximation technique, where data is projected into lower-dimensional subspaces of RKHS (spanned by so called \emph{inducing variables}). Such an idea is popular in fitting sparse approximations to Gaussian process (GP) regression models, allowing reduction of the computational cost from $\mathcal{O}(m^3)$ to $\mathcal{O}(n^2m)$ where $n \ll m$ is the number of inducing variables. To the best of our knowledge, Nystr\"om approximation was not studied in the context of hypothesis testing. Random Fourier feature (RFF) approximations \cite{ali2007}, however, due to their relationship with evaluations of empirical characteristic functions, do have a rich history in the context of statistical testing -- as discussed in \cite{ChwRamSejGre2015}, which also proposes an approach to scale up kernel-based two sample tests by additional smoothing of characteristic functions, thereby improving the test power and its theoretical properties. Moreover, the approximation strategy of MMD and two-sample testing through primal representation using RFF have also been studied in \cite{Zhao2015,ErrorRFF,David_thesis}. In addition, \cite{Lopez2013} first proposed the idea of applying RFF in order to construct an approximation to a kernel-based dependence measure. More specifically, they develop Randomised Canonical Correlation Analysis (RCCA) (see also \cite{Lopez2014,David_thesis}) approximating the non-linear kernel-based generalisation of the Canonical Correlation Analysis \cite{Lai_KCCA,Bach_KCCA} and using a further copula transformation, construct a test statistic termed RDC (randomised dependence coefficient) requiring $O(m\log m)$ time to compute. Under suitable assumptions, Bartlett's approximation \cite{Mardia_RCCA} provides a closed form expression for the asymptotic null distribution of this statistic which further results in a  distribution free test, leading to an attractive option for large-scale independence testing. We extend these ideas based on RFF to construct approximations of HSIC and dCov/dCor, which are conceptually distinct kernel-based dependence measures from that of kernel CCA, i.e., they measure different types of norms of RKHS operators (operator norm vs Hilbert-Schmidt norm). 

In fact, the Nystr\"om and RFF approximations can also be viewed through the lense of nonlinear canonical analysis framework introduced by \cite{Dauxois_nonlinear_indeptest}. This is the earliest example we know where nonlinear dependence measures based on spectra of appropriate Hilbert space operators are studied. In particular, the cross-correlation operator with respect to a dictionary of basis functions in $L_2$ (e.g. B-splines) is considered in \cite{Dauxois_nonlinear_indeptest}. \cite{Huang2009} links this framework to the RKHS perspective. The functions of the spectra that were considered in \cite{Dauxois_nonlinear_indeptest} are very general, but the simplest one (sum of the squared singular values) can be recast as the NOrmalised Cross-Covariance Operator (NOCCO) of \cite{Fukumizu08kernelmeasures}, which considers the Hilbert-Schmidt norm of the cross-correlation operator on RKHSs and as such extends kernel CCA to consider the entire spectrum. While in this work we focus on HSIC (Hilbert-Schmidt norm of the \emph{cross-covariance operator}), which is arguably the most popular kernel dependence measure in the literature, a similar Nystr\"om or RFF approximation can be applied to NOCCO as well - we leave this as a topic for future work.

The paper is structured as follows: in Section \ref{sec:background}, we first provide some necessary definitions from the RKHS theory and review the aforementioned Hilbert-Schmidt Independence Criterion (HSIC) and discuss its biased and unbiased quadratic time estimators. 
Then, Section \ref{sec:nulldistribution} gives the asymptotic null distributions of estimators (proofs provided in Section \ref{sec:Appendix}). In Section \ref{sec:block}, we develop a block-based HSIC estimator and derive its asymptotic null distribution. Following this, a linear time asymptotic variance estimation approach is proposed. In Section \ref{sec:Nystrom} and \ref{sec:RFF}, we propose Nystr\"om HSIC and RFF HSIC estimator respectively, both with the corresponding linear time null distribution estimation approaches. Finally, in Section \ref{sec:experiments}, we explore the performance of the three testing approaches on a variety of challenging synthetic data. 

%for the asymptotic null distribution of the biased HSIC estimator since it lays the theoretical foundations for the Nystr\"om spectral approach (Section 4.2) and RFF spectral approach (Section 5.2). Section 2.2 considers different null distribution estimation approaches. 

%%%%%%%%%%%%%%%%%%%%%%%
%%%%%%%   Background      %%%%%%%
%%%%%%%%%%%%%%%%%%%%%%%

\section{Background}
\label{sec:background}
This section starts with a brief overview of the key concepts and notation required to understand the RKHS theory and kernel embeddings of probability distributions into the RKHS. It then provides the definition of HSIC which will serve as a basis for later independence tests. We review the quadratic time biased and unbiased estimators of HSIC as well as their respective asymptotic null distributions. As the final part of this section, we outline the construction of independence tests in quadratic time. \\

\subsection{RKHS and Embeddings of Measures}

Let $\mathcal{Z }$ be any topological space on which Borel measures can be defined. By $\mathcal{M(Z)}$ we denote the set of all finite signed Borel measures on $\mathcal{Z}$ and by $\mathcal{M}^1_+ (\mathcal{Z})$ the set of all Borel probability measures on $\mathcal{Z}$. We will now review the basic concepts of RKHS and kernel embeddings of probability measures. For further details, see \cite{BerTho04, Steinwart2008book,Bharath}.
%Let $P_X \in \mathcal{M}^1_+ (\mathcal{X})$ and $P_Y \in \mathcal{M}^1_+ (\mathcal{Y})$ have finite first moment. 

%RKHS
\begin{definition} \label{def: Reproducing}
Let $\mathcal{H}$ be a Hilbert space of real-valued function defined on $\mathcal{Z}$. A function $k:\mathcal{Z} \times \mathcal{Z} \rightarrow \mathbb{R}$ is called a {\bf reproducing kernel} of $\mathcal{H}$ if: 
\begin{enumerate}
\item $\forall z \in \mathcal{Z}, k(\cdot,z) \in \mathcal{H}$
\item $\forall z \in \mathcal{Z}, \forall f \in \mathcal{H}, \langle f,k(\cdot,z) \rangle_\mathcal{H} = f(z).$
\end{enumerate}
If $\mathcal{H}$ has a reproducing kernel, it is called a {\bf Reproducing Kernel Hilbert Space} (RKHS). 
\end{definition}

As a direct consequence, for any $x,y \in \mathcal{Z}$, 
\begin{equation}
\label{eq:canonical_feature_map}
k(x,y) = \langle k(\cdot,x), k(\cdot,y) \rangle_\mathcal{H}.
\end{equation}
In machine learning literature, a notion of \emph{kernel} is understood as an inner product between feature maps \cite{Steinwart2008book}. By \eqref{eq:canonical_feature_map}, every reproducing kernel is a kernel in this sense, corresponding to a \emph{canonical feature map} $x\mapsto k(\cdot,x)$.

For $x,y \in \mathbb{R}^p$, some examples of reproducing kernels are
\begin{itemize}
\item Linear kernel: $k(x,y) = x^T y$; 
\item Polynomial kernel of degree $d \in \mathbb{N}$: $k(x,y) = (x^T y + 1)^d $;
\item Gaussian kernel with bandwidth $\sigma > 0$: $k(x,y) = \exp(-\frac{\| x- y\|^2}{2\sigma^2})$;
\item Fractional Brownian motion covariance kernel with parameter $H\in(0,1)$: $k(x,y)=\frac{1}{2}\left(\| x\|^{2H}+\| y\|^{2H}-\| x-y\|^{2H}\right)$
\end{itemize}

Checking whether a given function $k$ is a valid reproducing kernel can be onerous. Fortunately, the Moore-Aronszajn theorem \cite{Aron} gives a simple characterisation: for any symmetric, positive definite function $k: \mathcal{Z} \times \mathcal{Z} \rightarrow \mathbb{R}$, there exists a unique Hilbert space of functions $\mathcal{H}$ defined on $\mathcal{Z}$ such that $k$ is reproducing kernel of $\mathcal{H}$ \cite{BerTho04}. RKHS are precisely the space of functions where norm convergence implies pointwise convergence and are as a consequence relatively well behaved comparing to other Hilbert spaces. In nonparametric testing, as we consider here, a particularly useful setup will be representing probability distributions and, more broadly, finite signed Borel measures $\nu\in\mathcal{M(Z)}$ with elements of an RKHS \citep{probembedding}.\\

% RKHS need not be defined through reproducing kernels. It can also be defined in terms of evaluational functionals, and it is precisely the space of functions where norm convergence implies point-wise convergence. More details are given in  \cite{Steinwart2008book} Definition 4.18(ii). The proofs of existence and uniqueness of reproducing kernels can be found in \cite{lec}. In particular, Riesz representation theorem shows that $\forall z \in \mathcal{Z}$ there exists a unique element $k(.,\mathcal{Z})$ in $\mathcal{H}_k$ that satisfies Definition \ref{def: Reproducing}(2) \cite{lec}. 

%kernel embedding 
\begin{definition}
Let $k$ be a kernel on $\mathcal{Z}$, and $\nu \in \mathcal{M(Z)}$. The {\bf kernel embedding } of measure $\nu$ into the RKHS $\mathcal{H}_k$ is $\mu_k (\nu) \in  \mathcal{H}_k$ such that $\int f(z) d\nu(z) = \  \langle f,\mu_k(\nu) \rangle _{\mathcal{H}_k} \forall f \in \mathcal{H}_k.$  
\end{definition}

It is understood from this definition that the integral of any RKHS function $f$ under the measure $\nu$ can be evaluated as the inner product between $f$ and the kernel embedding $\mu_k (\nu)$  in the RKHS $ \mathcal{H}_k.$ As an alternative, the kernel embedding can be defined through the use of Bochner integral $\mu_k (\nu) = \int k(\cdot,z) d\nu(z)$. Any probability measure is mapped to the corresponding expectation of the canonical feature map. By Cauchy-Schwarz inequality and the Riesz representation theorem, a sufficient condition for the existence of an embedding of $\nu$ is that $\nu\in \mathcal{M}^{1/2}_{k}(\mathcal{Z})$, where we adopt notation from \cite{SejSriGreFuku13}:
$\mathcal{M}^\theta_{k}(\mathcal{Z}) = \left \{ \nu \in \mathcal{M}(Z): \int k^\theta(z,z) d|\nu|(z) < \infty  \right \}$,
which is, e.g. satisfied for all finite measures if $k$ is a bounded function (such as Gaussian kernel).

Embeddings allow measuring distances between probability measures, giving rise to the notion of Maximum Mean Discrepancy (MMD) \cite{Borgwardt2006,Gretton2sample}.
%MMD
\begin{definition}
 Let $k$ be a kernel on $\mathcal{Z}$. The squared distance between the kernel embeddings of two probability measures $P$ and $Q$ in the RKHS, $\text{MMD}_k(P,Q)=\|\mu_k(P)-\mu_k(Q)\|_{\mathcal H_k}^2$ is called {\bf Maximum Mean Discrepancy (MMD)} between $P$ and $Q$ with respect to $k$.
\end{definition}

When the corresponding kernels are \emph{characteristic} \cite{Bharath}, embedding is injective and MMD is a metric on probability measures. The estimators of MMD are useful statistics in nonparametric two-sample testing \cite{Gretton2sample}, i.e. testing if two given samples are drawn from the same probability distribution. For any kernels $k_{\mathcal X}$ and $k_{\mathcal Y}$ on the respective domains $\mathcal X$ and $\mathcal Y$, it is easy to check that $k=k_{\mathcal X}\otimes k_{\mathcal Y}$ given by 
\begin{equation}
 k\left(\left(x,y\right),\left(x',y'\right)\right)=k_{\mathcal X}(x,x')k_{\mathcal Y}(y,y')
\end{equation}
is a valid kernel on the product domain $\mathcal X \times \mathcal Y$. Its canonical feature map is $(x,y)\mapsto k_{\mathcal X}(\cdot,x)\otimes k_{\mathcal Y}(\cdot,y)$ where $\varphi_{x,y}=k_{\mathcal X}(\cdot,x)\otimes k_{\mathcal Y}(\cdot,y)$ is understood as a function on $\mathcal X \times \mathcal Y$, i.e. $\varphi_{x,y}(x',y')=k_{\mathcal X}(x',x)k_{\mathcal Y}(y',y)$. The RKHS of $k=k_{\mathcal X}\otimes k_{\mathcal Y}$ is in fact isometric to $\mathcal H_{k_\mathcal X}\otimes \mathcal H_{k_\mathcal Y}$, which can be viewed as the space of Hilbert-Schmidt operators between $\mathcal H_{k_\mathcal Y}$ and $\mathcal H_{k_\mathcal X}$ (Lemma 4.6 of \cite{Steinwart2008book}). We are now ready to define an RKHS-based measure of dependence between random variables $X$ and $Y$. 

\begin{definition}
 Let $X$ and $Y$ be random variables on domains $\mathcal X$ and $\mathcal Y$ (non-empty topological spaces). Let $k_{\mathcal X}$ and $k_{\mathcal Y}$ be kernels on $\mathcal X$ and $\mathcal Y$ respectively. {\bf Hilbert-Schmidt Independence Criterion (HSIC)} $\Xi_{k_\mathcal X,k_\mathcal Y}(X,Y)$ of $X$ and $Y$ is MMD between the joint measure $P_{XY}$ and the product of marginals $P_{X}P_{Y}$, computed with the product kernel $k=k_{\mathcal X}\otimes k_{\mathcal Y}$, i.e.,
 \begin{equation}
  \Xi_{k_\mathcal X,k_\mathcal Y}(X,Y)=\left\| \mathbb{E}_{XY}[k_\mathcal{X}(.,X) \otimes k_\mathcal{Y}(.,Y)] - \mathbb{E}_X k_\mathcal{X}(.,X) \otimes \mathbb{E}_Y k_\mathcal{Y}(.,Y) \right\|^2_{\mathcal{H}_{k_\mathcal{X}\otimes k_\mathcal{Y}}}.
 \end{equation}

\end{definition}

 HSIC is well defined whenever $P_X \in \mathcal{M}^1_{k_{\mathcal X}}(\mathcal{X})$ and $P_Y \in \mathcal{M}^1_{k_{\mathcal Y}}(\mathcal{Y})$ as this implies $P_{XY} \in \mathcal{M}^{1/2}_{k_{\mathcal X}\otimes k_{\mathcal Y}}(\mathcal{X} \times \mathcal{Y})$ \cite{SejSriGreFuku13}. The name of HSIC comes from the operator view of the RKHS $\mathcal{H}_{k_\mathcal{X}\otimes k_\mathcal{Y}}$. Namely, the difference between embeddings $\mathbb{E}_{XY}[k_\mathcal{X}(.,X) \otimes k_\mathcal{Y}(.,Y)] - \mathbb{E}_X k_\mathcal{X}(.,X) \otimes \mathbb{E}_Y k_\mathcal{Y}(.,Y)$ can be identified with the cross-covariance operator $C_{XY}:\mathcal H_{k_\mathcal Y}\to\mathcal H_{k_\mathcal X}$ for which $\langle f,C_{XY}g\rangle_{\mathcal H_{k_\mathcal X}}=\text{Cov}\left[f(X)g(Y)\right]$, $\forall f\in\mathcal H_{k_\mathcal X},g\in\mathcal H_{k_\mathcal Y}$ \cite{gretbousmol2005,greker08}. HSIC is then simply the squared Hilbert-Schmidt norm $\Vert C_{XY}\Vert_{HS}^2$ of this operator, while distance correlation (dCor) of \cite{Cordis2007,SR2009} can be cast as $\Vert C_{XY}\Vert_{HS}^2 / \Vert C_{XX}\Vert_{HS}\Vert C_{YY}\Vert_{HS}$ \cite[Appendix A]{SejSriGreFuku13}. In the sequel, we will suppress dependence on kernels $k_\mathcal X$ and $k_\mathcal Y$ in notation $\Xi_{k_\mathcal X,k_\mathcal Y}(X,Y)$ where there is no ambiguity. \\

Repeated application of the reproducing property gives the following equivalent representation of HSIC \cite{probembedding}:
\begin{proposition}
\label{prop:HSIC_as_expectations}
The HSIC of $X$ and $Y$ can be written as:
\begin{align}
\label{eq:HSIC_as_expectations}
\Xi(X,Y)&= \mathbb{E}_{XY} \mathbb{E}_{X'Y'} k_\mathcal{X}(X,X') k_\mathcal{Y}(Y,Y') \\\nonumber
& \qquad+ \mathbb{E}_{X} \mathbb{E}_{X'} k_\mathcal{X}(X,X')  \mathbb{E}_{Y} \mathbb{E}_{Y'} k_\mathcal{Y}(Y,Y') \\\nonumber
& \qquad\qquad- 2 \mathbb{E}_{X'Y'}[\mathbb{E}_X k_\mathcal{X}(X,X') \mathbb{E}_Y k_\mathcal{Y}(Y,Y')]. 
\end{align}
\end{proposition}

% %\text{MMD}_k (P_{XY},P_XP_Y) \\
%  &= \| \mathbb{E}_{XY}[k_\mathcal{X}(.,X) \otimes k_\mathcal{Y}(.,Y)] - \mathbb{E}_X k_\mathcal{X}(.,X) \otimes \mathbb{E}_Y k_\mathcal{Y}(.,Y) \|^2_{\mathcal{H}_{k_\mathcal{X}} \otimes \mathcal{H}_{k_\mathcal{Y}}} \\
% & \quad 

\subsection{Estimation of HSIC}

Using the form of HSIC in \eqref{eq:HSIC_as_expectations}, given an iid sample of \textbf{z} = $\{ (x_i, y_i) \}^m_{i=1}$ from the joint distribution $P_{XY}$, an unbiased estimator of HSIC can be obtained as a sum of three U-statistics \cite{greker08}:

\begin{align}
\label{eq:HSICU}
\Xi_u(\mathbf{z}) &= \frac{(m-2)!}{m!} \sum_{(i,j) \in \mathbf{i}_2^m} (K_x)_{ij} (K_y)_{ij} \\\nonumber
&\qquad+ \frac{(m-4)!}{m!} \sum_{(i,j,q,r) \in \mathbf{i}_4^m} (K_x)_{ij} (K_y)_{qr} \\\nonumber
 &\qquad\qquad- 2 \frac{(m-3)!}{m!} \sum_{(i,j,q) \in \mathbf{i}_3^m} (K_x)_{ij} (K_y)_{iq},
\end{align}
where the index set $\mathbf{i}^m_r$ denotes the set of all $r$-tuples drawn without replacement from $\{ 1, ... , m \}$, $(K_x)_{ij} := k_{\mathcal X}(x_i,x_j)$ and $(K_y)_{ij} := k_{\mathcal Y}(y_i,y_j)$.  Na\"ive computation of \eqref{eq:HSICU} would require $\mathcal O(m^4)$ operations. However, an equivalent form which needs $\mathcal O(m^2)$ operations is given in \cite{SonSmoGreBedetal12} as 
\begin{equation}
\Xi_u(\mathbf{z}) = \frac{1}{m(m-3)} \bigg [ \text{tr}(\tilde{K}_x \tilde{K}_y) + \frac{\mathds{1}^T \tilde{K}_x \mathds{1} \mathds{1}^T \tilde{K}_y \mathds{1}}{(m-1)(m-2)} - \frac{2}{m-2} \mathds{1}^T \tilde{K}_x \tilde{K}_y \mathds{1} \bigg]
\end{equation}
where $\tilde{K}_x = K_x - diag(K_x)$ (i.e. the kernel matrix with diagonal elements set to zero) and similarly for $\tilde{K}_y$. $\mathds{1}$ is a vector of 1s of relevant dimension.\\

We will refer to the above as the quadratic time estimator. % actually a navie computation leads to m^4.
\cite{greker08} note that the $V$-statistic estimator (or the quadratic time biased estimator) of HSIC can be an easier-to-use alternative for the purposes of independence testing, since the bias is accounted for in the asymptotic null distribution. The $V$-statistic is given by 
\begin{align*}
\Xi_b(\mathbf{z}) &= \frac{1}{m^2} \sum_{i,j}^m (K_x)_{ij} (K_y)_{ij}+ \frac{1}{m^4} \sum_{i,j,q,r}^m (K_x)_{ij} (K_y)_{qr} - 2 \frac{1}{m^3} \sum_{i,j,q}^m (K_x)_{ij} (K_y)_{iq}, 
\end{align*}
where the summation indices are now drawn with replacement. Further, it can be simplified as follows to reduce the computation:
\begin{align}
\Xi_b(\mathbf{z})  &= \frac{1}{m^2}\text{Trace}(K_x H K_y H) = \frac{1}{m^2}\langle HK_xH, HK_yH \rangle \label{eq:HSIC2}
\end{align}
where $H = I_m- \frac{1}{m}\mathds{1}\mathds{1}^T$ is an $m \times m$ centering matrix. \eqref{eq:HSIC2} gives an intuitive understanding of the HSIC statistic:  it measures average similarity between the centered kernel matrices, which are in turn similarity patterns within the samples.\footnote{A straightforward estimator of dCor \cite{Cordis2007,SR2009} is then given by normalising $\Xi_b(\mathbf{z})$ by the Frobenius norms of $HK_xH$ and $HK_yH$, i.e.,  $\widehat{dCor}({\bf z}) = \frac{\langle HK_xH, HK_yH \rangle}{\Vert HK_xH\Vert_F\Vert HK_yH\Vert_F}$}

%This following part is only true if we have a finite dimensional feature space, in which case, taking transpose make sense. 
%If we let $\Phi \Phi^T$ be the explicit feature representation of $K_x$ and $\Psi \Psi^T$ be that of $K_y$, the above expression can be further simplified as 
%\begin{align}
 %&\hat \gamma^2_k(P_{XY},P_XP_Y) \\
 %&= \langle \frac{1}{m} (H\Phi)^TH\Psi, \frac{1}{m}(H\Phi)^TH\Psi \rangle \\
 %&= \frac{1}{m^2} Trace((H\Phi)^TH\Psi \Psi^T H H \Phi) \\
 %&= \frac{1}{m^2} Trace(\Phi \Phi^TH H\Psi \Psi^T H H ) \\
 %&=\frac{1}{m^2} Trace(K_xHK_y H )
%\end{align}
%We hence obtain the expression in \eqref{eq:HSIC}.

\subsection{Asymptotic Null Distribution of Estimators} \label{sec:nulldistribution}
The asymptotic null distribution of the biased HSIC statistic defined in \eqref{eq:HSIC2} computed using a given data set converges in distribution in Theorem \ref{th:asynull} below. This asymptotic distribution builds the theoretical foundation for the spectral testing approach (described in Section \ref{Sec: Spectral}) that we will use throughout the paper.

% Asymptotic Null Distribution of Biased HSIC
\begin{theorem}
\label{th:asynull} 
($\mathbf{Asymptotic \  Null \ Distribution \ of \ the \ Biased \ HSIC }$)
Under the null hypothesis, let the dataset $\mathbf{z} = \{(x_i,y_i)\}^m_{i=1} \iid P_{XY}=P_XP_Y$, with $P_X \in \mathcal{M}^2_{k_{\mathcal X}}(\mathcal{X})$ and $P_Y \in \mathcal{M}^2_{k_{\mathcal Y}}(\mathcal{Y})$, then 
\begin{equation}
\label{eq:HSICasy}
m \Xi_{b,k_\mathcal{X},k_\mathcal{Y}}(\mathbf{Z}) \xrightarrow{D} \sum^\infty_{i=1} \sum^\infty_{j=1} \lambda_i \eta_j N^2_{i,j}
\end{equation}
where $N_{i,j} \iid \mathcal{N}$(0,1), $ \forall \ i,j \in \mathbb{N}$ and $\{\lambda_i\}^\infty_{i=1}$, $\{\eta_j\}^\infty_{j=1}$ are the eigenvalues of the integral kernel operators $S_{\tilde k_{P_x}}$ and $S_{\tilde k_{P_y}}$, where 
the integral kernel operator $S_{\tilde k_{P}}: L^2_P(\mathcal{Z}) \rightarrow  L^2_P(\mathcal{Z})$ is given by 
\begin{equation}
S_{\tilde k_P} g(z) = \int_{\mathcal{Z}} \tilde k_P(z,w)g(w) dP(w).
\end{equation} 
where $\tilde k_P(z, z')$ is the kernel centred at probability measure $P$: % \in \mathcal{M}^1_+ (\mathcal{Z})$: 
\begin{align} 
\tilde k_P (z,z'):&= \langle k(z,.) - \mathbb{E}_Wk(W, .), k(z',.) - \mathbb{E}_Wk(W, .) \rangle \nonumber \\
&= k(z,z') + \mathbb{E}_{WW'} k(W,W') - \mathbb{E}_W k(z,W) - \mathbb{E}_W k(z',W) \label{eq: centred kernel},
 \end{align}
with $W,W'\iid P.$
\end{theorem}
For completeness, the proof of this theorem, which is a consequence of \cite[Theorem 2.7]{L2013} and the equivalence between distance-based and RKHS-based independence statistics \cite{SejSriGreFuku13} is given in the Appendix \ref{sec:Appendix}. As remarked by \cite{SejSriGreFuku13}, the finite marginal moment conditions imply that the integral operators $S_{\tilde k_{\mathcal X}}$ and $S_{\tilde k_{\mathcal Y}}$ are trace class and hence Hilbert-Schmidt \cite{RS1980}. Anderson et al. noted that the form of the asymptotic distribution of the V statistics requires the integral operators being trace class but that of the U statistics only requires them being Hilbert-Schmidt \cite{Anderson1994,SejSriGreFuku13}. Using the same notation as in the case of the V-statistic, the asymptotic distribution of the U-statistic in \eqref{eq:HSICU} can be written as:
\begin{equation}
\label{eq:HISCdis}
m \Xi_{u,k_\mathcal{X},k_\mathcal{Y}}(\mathbf{Z}) \xrightarrow{D} \sum^\infty_{i=1} \sum^\infty_{j=1} \lambda_i \eta_j (N_{i,j}^2-1)
\end{equation}
under the null hypothesis. 

We note that \cite{KacperHSIC} (Lemma 2 and Theorem 1) proves a more general result, applicable to dependent observations under certain mixing conditions where the i.i.d. setting is a special case. Moreover, \citep{Paul3varInter} (Theorem 5 and 6) provides another elegant proof in the context of three-variable interaction testing from \cite{SejGreBer2013x}. However, both \cite{KacperHSIC} and \citep{Paul3varInter} assume boundedness of $k_\mathcal{X}$ and $k_\mathcal{Y}$, while our proof in the Appendix \ref{sec:Appendix} assumes a weaker condition of finite second moments for both $k_\mathcal{X}$ and $k_\mathcal{Y}$, thus making the result applicable to unbounded kernels such as the Brownian motion covariance kernel.

Under the alternative hypothesis that $P_X P_Y \not = P_{XY}$, \cite{greker08} remarked that $m \Xi_{b,k_\mathcal{X},k_\mathcal{Y}}(\mathbf{Z})$ converges to $HSIC$ with the corresponding appropriately centred and scaled Gaussian distribution as $m \rightarrow \infty $:
\begin{equation}
\label{eq:alterasy}
\sqrt{m} ( \Xi_{b,k_\mathcal{X},k_\mathcal{Y}}(\mathbf{Z}) - HSIC) \xrightarrow{D} \mathcal{N}(0, \sigma^2_u)
\end{equation}
where the variance $\sigma^2_u = 16(\mathbb{E}_{z_i} (\mathbb{E}_{z_j,z_q,z_r}(h_{ijqr}))^2  -HSIC)$ and $h_{ijqr}$ is defined as 
\begin{equation}
h_{ijqr} = \frac{1}{4!} \sum^{(i,j,q,r)}_{(t,u,v,w)} (K_x)_{tu} (K_y)_{tu} + (K_x)_{tu} (K_y)_{vw} + (K_x)_{tu} (K_y)_{tv}
\end{equation}
with all ordered quadruples $(t,u,v,w)$ drawn without replacement from $(i,j,q,r)$ and assuming $\mathbb{E} (h^2) < \infty$. In fact, under the alternative hypothesis, the difference between $m \Xi_{b}(\mathbf{Z})$ (i.e. the V-statistic) and the U-statistic drops as $1/m$ and hence asymptotically the two statistics converges to the same distribution \cite{greker08}. \\

\subsection{Quadratic Time Null Distribution Estimations}
We would like to design independence tests with an asymptotic Type I error of $\alpha$ and hence we need an estimate of the $(1-\alpha)$ quantile of the null distribution. Here, we consider two frequently used approaches, namely the permutation approach and the spectral approach, that require at least quadratic time both in terms of memory and computation time. The biased V-statistic will be used because of its neat and compact formulation.

\subsubsection{Permutation Approach}
\label{Sec:Permutation}
Consider an iid sample $\mathbf{z} = \left\{ (x_i,y_i)\right \}^m_{i=1}$ with chosen kernels $k_{\mathcal X}$ and $k_{\mathcal Y}$ respectively, the permutation/bootstrap approach \citep{arcones1992} proceed in the following manner. Suppose the total number of shuffles is fixed at $N_p$, we first compute $\Xi_{k_\mathcal{X},k_\mathcal{Y}}(\mathbf{z})$ using $\mathbf{z}$, $k_{\mathcal X}$ and $k_{\mathcal Y}$. Then, for each shuffle, we fix the $\{ x_i \}^m_{i=1}$ and randomly permute the $\{ y_i \}^m_{i=1}$ to obtain $\mathbf{z}^* = \left\{ (x_i,y^*_i)\right \}^m_{i=1}$ and subsequently compute $\Xi^*_{k_\mathcal{X},k_\mathcal{Y}}(\mathbf{z^*})$. The one-sided p-value in this instance is the proportion of HSIC values computed on the permuted data that is greater than or equal to $\Xi_{k_\mathcal{X},k_\mathcal{Y}}(\mathbf{z})$. \\

The computational time is $\mathcal{O}$(number of shuffles $\times m^2)$ for this approach, where the number of shuffles determines the extend to which we have explored the sampling distribution. In other words, a small number of shuffles means that we may only obtained realisations from the mode of the distribution and hence the tail structure is not adequately captured. Although a larger number of shuffles ensures the proper exploration of the sampling distribution, the computation cost can be high. 

%Note{With a permutation test, the shuffled data set should look the same as the real data set (i.e. the data generated by the sampling distribution). How I understand this is that the shuffled data set can be regarded as realisations from the sampling distribution?  If the null is not true, then the shuffled data set looks different from the real data.} 

\subsubsection{Spectral Approach}
\label{Sec: Spectral}

Gretton et al. has shown (Theorem 1 \cite{GreFukHarSri09}) that the empirical finite sample estimate of the null distribution converges in distribution to its population counterpart provided that the eigenspectrums $\{ \gamma_r \}^\infty_{r=1}$ of the integral operator  $S_{\tilde k }$: $L^2_\theta (\mathcal{X} \times \mathcal{Y}) \rightarrow L^2_\theta (\mathcal{X} \times \mathcal{Y})$ is square root summable, i.e. 
$$\sum^\infty_{r=1} \sqrt{ \gamma_r }= \sum^\infty_{i=1} \sum^\infty_{j=1} \sqrt{\lambda_i\eta_j} < \infty.$$
Note, the integral operator $S_{\tilde k }$ is the tensor product of the operators $S_{\tilde k_{\mathcal X} } $ and $S_{\tilde k_{\mathcal X} } $: 
$$
S_{\tilde k } g(x,y) = \int_{\mathcal{X} \times \mathcal{Y}}  \tilde k_\mu (x,x') \tilde k_\nu (y,y') g(x',y') d\theta(x',y')
$$
and the eigenvalues of this operator is hence the product of the eigenvalues of these two operators.  \\

The spectral approach \cite{GreFukHarSri09, ZhPeJanSch11} requires that we first calculate the centred Gram matrices $\widetilde{K}_X = HK_{X}H$  and  $\widetilde{K}_Y = HK_{Y}H$ for the chosen kernel $k_{\mathcal X}$ and $k_{\mathcal Y}$. Then, we compute the $m \Xi_{b,k_\mathcal{X},k_\mathcal{Y}}(\mathbf{z})$ statistics according to \eqref{eq:HSIC2}. Next, the spectrums (i.e. eigenvalues) $\{\lambda_i\}^m_{i=1}$ and $\{\eta_i\}^m_{i=1}$ of $\widetilde{K}_X$ and $\widetilde{K}_Y$ are respectively calculated. The empirical null distribution can be simulated by simulating a large enough i.i.d samples from the standard Normal distribution \citep{ZhPeJanSch11} and then generate the test statistic according to \eqref{eq:HSICasy}. Finally, the p-value is computed by calculating the proportion of simulated samples that are greater than or equal to the observed $m \Xi_{b,k_\mathcal{X},k_\mathcal{Y}}(\mathbf{z})$ value.

Additionally, \cite{ZhPeJanSch11} has provided an approximation to the null distribution with a two-parameter Gamma distribution. Despite the computational advantage of such an approach, the permutation and spectral approaches are still preferred since there is no consistency guarantee for the Gamma distribution approach. % and hence it can sometimes hinder the test performance.

%%%%%%%%%%%%%%%%%%%%%%%%%
%%%%%%%%     Block HSIC      %%%%%%%%
%%%%%%%%%%%%%%%%%%%%%%%%%
\section{Block-based HSIC} \label{sec:block}
The quadratic time test statistics are prohibitive for large dataset as it requires $\mathcal{O}(m^2)$ time in terms of storage and computation. Furthermore, one requires an approximation of the asymptotic null distribution in order to compute the p-value. As we discussed in the previous section, this is usually done by randomly permute the $Y$ observations (i.e. the permutation approach) or by performing an eigen-decomposition of the centred kernel matrices for $X$ and $Y$ (i.e. the spectral approach). Both approaches are expensive in terms of memory and can be computationally infeasible.  In this section, we propose a block-based estimator of HSIC which reduce the computational time to linear in the number of samples. The asymptotic null distribution of this estimator will be shown to have a simple form as a result of the Central Limit Theorem (CLT).

\subsection{The Block HSIC Statistic}
Let us consider that the sample is split into blocks of size $B\ll m$:
$\left\{ x_{i},y_{i}\right\}_{i=1}^{m} \iid P_{XY}$ becomes $\left\{ \left\{ x_{i}^{(b)}, y_{i}^{(b)} \right\} _{i=1}^{B}\right\} _{b=1}^{m/B}$ (where we assumed for simplicity that $m$ is divisible by $B$). We follow the approach from \cite{ZarGreBla13,BigHyp} and extend it to independence testing. We compute the unbiased HSIC statistics (Eq. \ref{eq:HSICU}) on each block $b \in \{ 1, ... , \frac{m}{B} \}$: 
\begin{equation}\label{eq:btest}
\hat \eta_{b} = \frac{1}{B(B-3)} \bigg [ \text{tr}(\tilde{K}_x^{(b)} \tilde{K}_y^{(b)}) + \frac{\mathds{1}^T \tilde{K}_x^{(b)} \mathds{1} \mathds{1}^T \tilde{K}_y^{(b)} \mathds{1}}{(B-1)(B-2)} - \frac{2}{B-2} \mathds{1}^T \tilde{K}_x^{(b)} \tilde{K}_y^{(b)} \mathds{1} \bigg]
\end{equation}
and average them over blocks to establish the block-based estimator for HSIC:
\begin{equation}
\hat \Xi_{k_\mathcal{X},k_\mathcal{Y}} = \frac{B}{m}\sum^{m/B}_{b=1} \hat \eta_{b}.  
\end{equation}

\subsection{Null Distribution of Block-Based HSIC }
For the block HSIC statistic, the asymptotic null distribution is a consequence of the Central Limit Theorem (CLT) under the regime where $m \rightarrow \infty$, $B \rightarrow \infty$ and $\frac{m}{B} \rightarrow \infty$\footnote{For example, $B=m^\delta \ \text{with } \delta \in (0,1)$}. First of all, note that the linear time test statistic $\hat \eta_k$ is an average of block-based statistics $\hat \eta_b$ for $b \in \{1, ..., \frac{m}{B}\}$ which are independent and identically distributed. Secondly, we recall that $\mathbb{E}(\hat \eta_b) = 0$ for $\hat \eta_b$ is an unbiased estimator of HSIC. Finally, $\mathbb{V}ar(\hat \eta_k) = \frac{B}{m} \mathbb{V}ar(\hat \eta_b) = \frac{B}{m}\frac{1}{B^2} \mathbb{V}ar(W)$ with $W$ being the random variable distributed according to $\sum^\infty_{i=1} \sum^\infty_{j=1} \lambda_i \eta_j (N_{i,j}^2-1)$. In the limit as $m \rightarrow \infty$, $B \rightarrow \infty$ and $\frac{m}{B} \rightarrow \infty$:
\begin{equation}
\label{ex: DisBtest}
\sqrt{mB} \hat \Xi_{k_\mathcal{X},k_\mathcal{Y}} \xrightarrow{D} \mathcal{N}(0, \sigma^2_{k,0}). 
\end{equation}
where the variance $\sigma^2_{k,0}$ is the variance of the null distributions in Expression (\ref{eq:HSICasy}) and (\ref{eq:HISCdis}) i.e. the variance of $W$ and it is given by 
\begin{align}
\sigma^2_{k,0} & = 2  \sum_{i=1}^\infty \sum_{j=1}^\infty \lambda_i^2 \eta_j^2 \\
                        & = 2\mathbb{E}_{XX'}(\tilde k^2_{P_X}(X,X'))\mathbb{E}_{YY'}(\tilde k^2_{P_Y}(Y,Y')) \label{eq:var}
\end{align}

\subsection{Linear Time Null Distribution Estimation}
Expression \eqref{ex: DisBtest} guarantees the Gaussianity of the null distribution of the block-based statistic and henceforth makes the computation of p-value straight-forward.  We simply return the test statistic $\sqrt{mB}\frac{ \hat \eta_{b}}{\sqrt{\hat \sigma^2_{k,0}}}$ and compare against the corresponding quantile of $\mathcal{N}(0,1 )$ which is the approach taken in \cite{optikernel12, ZarGreBla13, BigHyp}. Note that the resulting null distribution is actually a t-distribution but with a very large number of degrees of freedom, which can be treated as a Gaussian distribution. \\

The difficulty of estimating the null distribution lies in estimating $\sigma^2_{k,0}$. We suggest two ways to estimate such variance \citep{BigHyp}: within-block permutation and within-block direct estimation. These two approaches are at most quadratic in $B$ within each block which means that the computational cost of estimating the variance is of the same order as that of computing the statistic itself.  \\

Within-block permutation can be done as follows. Within each block, we compute the test statistic using \eqref{eq:btest}. At the same time, we track in parallel a sequence $\hat \eta^*_{b}$ obtained using the same formula but with $\{y_i\}^m_{i=1}$ underwent a permutation. The former is used to calculate the overall block statistics and the latter is used to estimate the null variance $\hat \sigma^2_{k,0} = B^2 \mathbb{V}ar \bigg [ \{\hat \eta^*_{b} \}^{m/B}_{b=1} \bigg ]$ as the independence between the samples holds by construction.  \\

Within block direct estimation can be achieved by using \eqref{eq:var} and the corresponding unbiased estimates of $\mathbb{E}_{XX'}(\tilde k^2_{P_X}(X,X'))$ and $\mathbb{E}_{YY'}(\tilde k^2_{P_Y}(Y,Y'))$ which can be calculated as follows. For $X$, the estimate of the variance for each block is given by \cite{SonSmoGreBedetal12}:
\begin{equation}
(\hat \sigma^2_{k,x})^{(b)} = \frac{2}{B(B-3)} \bigg[   \text{tr}(\tilde{K}^{(b)}_x \tilde{K}^{(b)}_x) + \frac{(\mathds{1}^T \tilde{K}^{(b)}_x \mathds{1})^2}{(B-1)(B-2)} - \frac{2}{B-2} \mathds{1}^T (\tilde{K}^{(b)}_x)^2  \mathds{1} \bigg]
\end{equation}
Then, we compute 
\begin{equation}
\hat \sigma^2_{k,x} = \frac{B}{m} \sum^{m/B}_{b=1} (\hat \sigma^2_{k,x})^{(b)}.
\end{equation}
to obtain an unbiased estimate for $\mathbb{E}_{XX'}(\tilde k^2_{P_X}(X,X'))$. Similarly, replacing all $x$ with $y$, we obtain an unbiased estimate for $\mathbb{E}_{YY'}(\tilde k^2_{P_Y}(Y,Y'))$. The estimate of the variance is therefore: 
\begin{equation}
\hat \sigma^2_{k,0} = 2 \hat \sigma^2_{k,x} \hat \sigma^2_{k,y} 
\end{equation}

%{Note that the variance of the statistic under the null now drops at
%a slower rate of $O(\frac{1}{nB})$ -- e.g., $O(\frac{1}{n^{1+\delta}})$,
%this is the price we pay for using the block-based statistic. In contrast,
%there is no loss in the convergence rate for the block-based statistic
%under the alternative. Variances of the asymptotic distributions in
%(\ref{eq: Basymp_H0}) and (\ref{eq: Basymp_HA}) are exactly the
%same as in (\ref{eq: asymp_H0}) and (\ref{eq: asymp_HA}) and do
%not depend on the rate of $B$. }

As remarked in \cite{BigHyp}, we note that under the null hypothesis, the approach undertaken by \cite{ZarGreBla13} is to estimate the null variance directly with the empirical variance of $\left\{ \hat{\eta}_{b}\right\} _{b=1}^{m/B}$. As the null variance is consistently estimated under the null hypothesis, this ensures the correct level of Type I error. However, without using the variance of the ``bootstrap samples", such an estimate of the null variance will systematically overestimate as $B$ grows with $m$. Hence, it will result in a reduced statistical power due to inflated p-values.

%%% From Remark 4 of Tech report for MMD %%%
%As an remark, we note that if the null holds, one can estimate variance under the null
%by looking at the empirical variance of $\left\{ \hat{\eta}_{k,b}\right\} _{b=1}^{m/B}$
%without the need for generating ``bootstrap samples" $\left\{ \hat{\eta}_{k,b}^{*}\right\} _{b=1}^{m/B}$.
%This is the approach undertaken in \cite{ZarGreBla13}. % In particular,
%empirical variance of $\left\{ \rho_{x}\rho_{y}B\hat{\eta}_{k,b}\right\} _{b=1}^{n/B}$
%will be approximately $\sigma_{k,0}^{2}$.  /////// not sure if this is still the case for HSIC 
%This ensures the correct
%level of the Type I error as the variance of the null is consistently
%estimated when the null holds. However, if the alternative holds,
%this is generally an incorrect estimate of the null variance.
%Namely, in this case the empirical variance of $\left\{ \rho_{x}\rho_{y}B\hat{\eta}_{k,b}\right\} _{b=1}^{n/B}$ grows with $B$ and will be approximately $\rho_{x}\rho_{y}B\sigma_{k,A}^{2}$.
%As $B$ grows with $m$, using such an estimate of the variance under the null will  
%systematically overestimate, and thus result
%in a reduced power as p-values tend to be inflated. 
%Therefore, we can expect this approach to drop in power as $m$
%increases. \\

Regarding the choice of $B$, \cite{ZarGreBla13} discussed that the null distribution is close to that guaranteed by the CLT when $B$ is small and hence the Type I error will be closer to the desired level. However, the disadvantage is the small statistical power for each given sample size. Conversely, \cite{ZarGreBla13} pointed out that a larger $B$ results in a lower variance empirical null distribution and hence higher power. Hence, they suggested a sensible family of heuristics is to set $B=\lfloor m^{\gamma}\rfloor$
for some $ 0<\gamma<1$. As a result, the complexity of the block-based test is $\mathcal{O}(Bm) = \mathcal{O}(m^{1+\gamma})$.

%%%%%%%%%%%%%%%%%%%%%%%%%%
%%%%%% Nystrom and RFF HSIC   %%%%%%%
%%%%%%%%%%%%%%%%%%%%%%%%%%

\section{Approximate HSIC through Primal Representations }
Having discussed how we can construct a linear time HSIC test by processing the dataset in blocks, we now move on to consider how the scaling up can be done through low rank approximations of the Gram matrix. In particular, we will discuss Nystr\"om type approximation (Section \ref{sec:Nystrom}) and Random Fourier Features (RFF) type approximation (Section \ref{sec:RFF}). Both types of approximation act directly on the primal representation of the kernel hence provide finite representations of the feature maps. 

%\dino{I would suggest moving this part up to null distributions to the section with primal representations}
Recall that the definition of HSIC of $X$ and $Y$ in \eqref{eq:HSIC_as_expectations} can also be written in terms of the cross covariance operator $C_{XY}$: $ \Xi(X,Y) = \| C_{XY} \|^2_{\mathcal{H}_{k_\mathcal{X}} \otimes \mathcal{H}_{k_\mathcal{Y}}}$ \citep{gretbousmol2005,greker08}.
Given a data set $\mathbf{z} = \{ (x_i,y_i) \}^m_{i=1}$ with $x_i \in \mathbb{R}^{D_x}$ and $y_i \in \mathbb{R}^{D_y}$ for all $i$, consider the empirical version of $\Xi_{k_\mathcal{X},k_\mathcal{Y}}(X,Y)$ with kernels $k_\mathcal{X}$ and $k_\mathcal{Y}$ for $X$ and $Y$ respectively:  
 \begin{align}
 &\hat \Xi_{k_\mathcal{X},k_\mathcal{Y}}(X,Y) \nonumber \\
  &= \left \Vert \frac{1}{m} \sum^{m}_{i = 1} k_\mathcal{X}(\cdot,x_i) \otimes k_\mathcal{Y}(\cdot,y_i) - \left ( \frac{1}{m} \sum^m_{i=1} k_\mathcal{X}(\cdot,x_i) \right )\otimes \left ( \frac{1}{m} \sum^m_{i=1} k_\mathcal{Y}(\cdot,y_i) \right ) \right \Vert ^2  \label{eq:NyHSIC}\\
 &= \left \Vert \frac{1}{m} \sum^{m}_{i = 1} \left( k_\mathcal{X}(\cdot,x_i)- \frac{1}{m} \sum^m_{r=1} k_\mathcal{X}(\cdot,x_r)  \right) \otimes \left ( k_\mathcal{Y}(\cdot,y_i)- \frac{1}{m} \sum^m_{r=1} k_\mathcal{Y}(\cdot,y_r)  \right )  \right \Vert^2%_{\mathcal{H}_{k_\mathcal{X}} \otimes \mathcal{H}_{k_\mathcal{Y}}}  
\end{align}
 %&= \left \Vert \frac{1}{m} \sum^{m}_{i = 1} k_\mathcal{X}(\cdot,x_i) \otimes k_\mathcal{Y}(\cdot,y_i) - \hat \mu_{k_\mathcal{X}} \otimes \hat \mu_{k_\mathcal{Y}} \right \Vert ^2_{\mathcal{H}_{k_\mathcal{X}} \otimes \mathcal{H}_{k_\mathcal{Y}}}  
%$ \hat \mu_{k_\mathcal{X}}$ and $\hat \mu_{k_\mathcal{Y}}$ are the empirical mean embeddings for $X$ and $Y$. 
where the Hilbert Schmidt norm is taken in the product space ${\mathcal{H}_{k_\mathcal{X}} \otimes \mathcal{H}_{k_\mathcal{Y}}}  $. Note, this empirical cross covariance operator is infinite dimensional. However, when approximate feature representations are used, the cross covariance operator is a finite dimensional matrix and hence the Hilbert-Schmidt norm is equivalent to the Frobenius norm ($F$). 

If we let $\bar \phi(x_i) = k_\mathcal{X}(\cdot,x_i)- \frac{1}{m} \sum^m_{r=1} k_\mathcal{X}(\cdot,x_r)$ and $\bar \psi(y_i) = k_\mathcal{Y}(\cdot,y_i)- \frac{1}{m} \sum^m_{r=1} k_\mathcal{Y}(\cdot,y_r)$, the above expression can be further simplified as
\begin{align}
&\hat \Xi_{k_\mathcal{X},k_\mathcal{Y}}(X,Y) \nonumber \\
&=\frac{1}{m^2} \sum^m_{i=1} \sum^m_{j=1} \langle \bar \phi(x_i) \otimes \bar \psi(y_i), \bar \phi(x_j) \otimes \bar \psi(y_j) \rangle \\
&=\frac{1}{m^2} \sum^m_{i=1} \sum^m_{j=1} \langle \bar \phi(x_i), \bar \phi(x_j) \rangle \langle  \bar \psi(y_i), \bar \psi(y_j) \rangle \\
%&= \frac{1}{m^2} Trace(\bar K_x \bar K_y) \\
&= \frac{1}{m^2} Trace(HK_xHHK_yH).
\end{align}
Hence, we obtain the expression in \eqref{eq:HSIC2}. If instead, we replace $\bar \phi(x_i)$ and $\bar \psi(y_i)$ by the corresponding low-rank approximations  $\tilde \phi(x_i) = \tilde k_\mathcal{X}(\cdot,x_i)- \frac{1}{m} \sum^m_{r=1} \tilde k_\mathcal{X}(\cdot,x_r)$ and $\tilde \psi(y_i) = \tilde k_\mathcal{Y}(\cdot,y_i)- \frac{1}{m} \sum^m_{r=1} \tilde k_\mathcal{Y}(\cdot,y_r)$, we can obtain the approximated HSIC statistics. The details of which are provided in the following sections.

%%%%%%%%%%%%%%%%%%%%%%
%%%%%%      Nystrom HSIC    %%%%%%
%%%%%%%%%%%%%%%%%%%%%%
\subsection{Nystr\"om HSIC}\label{sec:Nystrom}

In this section, we use the traditional Nystr\"om approach to provide an approximation that consider the similarities between the so called inducing variables and the given dataset. We will start with a review of Nystr\"om  method and then we will provide the explicit feature map representation of the Nystr\"om HSIC estimator. To finish, we will discuss two null distribution estimation approaches that cost at most linear in the number of samples. \\%$\mathcal{O}(n^2m)$ where $n$ is the number of inducing variables such that $n \ll m$. \\

The reduced-rank approximation matrix provided by Nystr\"om method \cite{Nystrom} represents each data point by a vector based on its kernel similarity to the inducing variables and the induced kernel matrix. %\cite{NvsRFF}.
 The approximation is achieved by randomly sample $n$ data points (i.e. inducing variables) from the given $m$ samples and compute the approximate kernel matrix $\tilde K \approx K$ as:  
\begin{equation}
\label{eq: Ny}
\tilde K = K_{m,n} K^{-1}_{n,n} K_{n,m}
\end{equation}
where each of $K_{r,s}$ can be think of as the $r \times s$ block of the full Gram matrix $K$ computed using all given samples. 
%The computational complexity of computing such statistic scales as $\mathcal{O}(n^3 + n^2 m)$.   \\
Further, we can write Eq. \ref{eq: Ny} as: 
\begin{equation}
\begin{aligned}
\tilde K &= K_{m,n} K^{-\frac{1}{2}}_{n,n} \left( K_{m,n} K^{-\frac{1}{2}}_{n,n}  \right)^{T} \\
 & = \tilde \Phi \tilde \Phi^T
\end{aligned}
\end{equation}
Hence, an explicit feature representation of $\tilde K$ is obtained. Note that \cite{Snelson06SG} further relaxed the setting and propose to use inducing points that are not necessarily a subset of the given data but only need to explain the dataset well for a good performance. 
%In \cite{Snelson06SG} (copied directly here) the author presented a Gaussian process (GP) regression model whose kernel covariance matrix is parameterised by the the locations of M pseudo-input points  Learning in the model involves finding a suitable setting of the parameters Ð an appropriate pseudo data set that explains the real data well.

\subsubsection{The Nystr\"om HSIC Statistic}
\label{Sec:Nystats}
To further reduce computation cost, we propose to approximate this reduced-rank kernel matrix $\tilde K$ with the uncentered covariance matrix $\tilde C$ that is $n \times n$: 
\begin{equation}
\begin{aligned}
\tilde C &=   \left( K_{m,n} K^{-\frac{1}{2}}_{n,n}  \right)^{T} K_{m,n} K^{-\frac{1}{2}}_{n,n}\\
 & = \tilde \Phi^T \tilde \Phi 
\end{aligned}
\end{equation}
Let us denote $\tilde C_X = \tilde \Phi_X^T \tilde \Phi_X $ and $\tilde C_Y = \tilde \Phi_Y^T \tilde \Phi_Y $. In order to approximate the biased HSIC estimator (Eq. \ref{eq:HSIC2}) using this explicit feature map representation, the $\tilde \Phi_X$ and $\tilde \Phi_Y$ needed to be centred. We suggest centre each column separately by subtracting its mean for both $\tilde \Phi_X$ and $\tilde \Phi_Y$, i.e. denote $ \hat \Phi_.=(I_m - \frac{1}{m}\mathds{1}\mathds{1}^T)\tilde \Phi_. = H \tilde \Phi. \in \mathcal{R}^{m \times n_.}$ for $X$ and $Y$ respectively. \\

Using the methods described above, we can substitute approximated kernel functions $\hat k_{\mathcal X} = \hat \Phi_X$ and $\hat k_{\mathcal Y} = \hat \Phi_Y$ into the empirical version of $\Xi_{k_\mathcal{X},k_\mathcal{Y}}(X,Y)$ \eqref{eq:NyHSIC}:
\begin{align} 
  &\hat \Xi_{Ny, \tilde k_\mathcal{X},\tilde k_\mathcal{Y}}(X,Y) \nonumber \\ %&= \left \Vert \frac{1}{m} \sum^{m}_{i = 1} \hat k_\mathcal{X}(\cdot,x_i) \otimes \hat k_\mathcal{Y}(\cdot,y_i) - \hat \mu_{\hat k_\mathcal{X}} \otimes \hat \mu_{\hat k_\mathcal{Y}} \right \Vert^2_{\mathcal{H}_{\hat k_\mathcal{X}} \otimes \mathcal{H}_{\hat k_\mathcal{Y}}}. \label{line1} \\
   & = \left \Vert \frac{1}{m} \sum^m_{i=1} \hat \Phi_X(x_i) \hat \Phi_Y(y_i)^T - \left ( \frac{1}{m} \sum^m_{i=1} \hat \Phi_X(x_i) \right ) \left (\frac{1}{m} \sum^m_{i=1} \hat \Phi_Y(y_i) \right)^T \right \Vert^2_F \\
   & = \left \Vert \frac{1}{m} \tilde \Phi_X^T  \tilde \Phi_Y \right \Vert^2_F 
 \end{align}
 where $\tilde \Phi_X(x_i) \in \mathbb{R}^{n_x}$ and $\tilde \Phi_Y(y_i) \in \mathbb{R}^{n_y}$ can both be computed in linear time in $m$. This is the biased Nystr\"om estimator of HSIC. Essentially, we approximate the cross covariance operator $C_{XY}$ by the Nystr\"om estimator $\tilde C_{XY} = \frac{1}{m} \tilde \Phi_X^T \tilde \Phi_Y \in \mathbb{R}^{n_x \times n_y}$ which only requires $\mathcal{O}(n_x n_y m )$.  In essence, the HSIC statistic computed using Nystr\"om as we described here is a HSIC statistic computed using a different kernel. As a remark, we note that it is not immediately clear how one can choose the inducing points optimally. For the synthetic data experiments in Section \ref{sec:experiments}, we simulate the inducing data from the same distribution as data $X$. But, we will leave the more general case as further work.

\subsubsection{Null Distribution Estimations}

Having introduced the biased Nystr\"om HSIC statistics, we will now move on to discuss two null distribution estimation methods, namely the permutation approach and the Nystr\"om spectral approach. The permutation approach is exactly the same as Section \ref{Sec:Permutation} with $\Xi_{k_\mathcal{X},k_\mathcal{Y}}(\mathbf{z})$ replaced by  $ \hat \Xi_{Ny, \tilde k_\mathcal{X},\tilde k_\mathcal{Y}}(\mathbf{z})$. It is worth noting that for each permutation, we need to simulate a new set of inducing points for $X$ and $Y$ such that $n_x, n_y  \ll m$ with $m$ being the number of samples. \\

Likewise, the Nystr\"om spectral approach is similar to that described in Section \ref{Sec: Spectral} where eigen-decompositions of the centred Gram matrices are required to simulate the null distribution. The difference is that we approximate the centred Gram matrices using  Nystr\"om method and the HSIC V-statistic is replaced by the Nystr\"om HSIC estimator $\hat \Xi_{Ny, \tilde k_\mathcal{X},\tilde k_\mathcal{Y}}(\mathbf{z})$.  So, the null distribution is then estimated using the eigenvalues from the covariance matrices $\tilde \Phi_X^T \tilde \Phi_X $ and $ \tilde \Phi_Y^T \tilde \Phi_Y $. In such a way, the computational complexity is reduced from the original $\mathcal{O}(m^3)$ to $\mathcal{O}(n_x^3+n_y^3+(n_x^2+n_y^2)m + n_xn_ym)$ i.e. linear in m.

%As eigen-decompositions for the centred Gram matrices are required to simulate the null distributions in spectral approach (Section \ref{Sec: Spectral}), it is particularly costly when the size of the dataset is large. We propose to use the covariance matrix $\Phi^\text{T} \Phi$ as an approximation to the Gram matrix using RFF expansions where each row of $\Phi$ is given by $z(x_j)$. However, note that the Gram matrices are centred, we suggest centring each column of $\Phi$ separately by subtracting the column mean. 

%%%%%%%%%%%%%%%%%%%%%%%%%%%%
%%%%%%%%%%   RFF HSIC %%%%%%%%%%%
%%%%%%%%%%%%%%%%%%%%%%%%%%%%

\subsection{Random Fourier Feature HSIC} \label{sec:RFF}

So far, we have looked at two large-scale approximation techniques that are applicable to any positive definite kernel. If the corresponding kernel also happens to be translation-invariant with the moment condition in \eqref{eq: momentcon}, however, an additional popular large-scale technique can be applied: random Fourier features of \cite{ali2007} which is based on Bochner's representation. In this section, we will first review Bochner's theorem and subsequently build up to how random Fourier features can be used to approximate large kernel matrices. Utilising it in the context of independence testing, we propose the novel RFF HSIC estimator and further consider two null distribution estimation approaches. 

\subsubsection{Bochner's Theorem}

Through the projection of data into lower dimensional randomised feature space, \cite{ali2007} proposed a method of converting the training and evaluation of any kernel machine into the corresponding operations of a linear machine. In particular, they showed using a randomised feature map $z : \mathcal{R}^d \rightarrow \mathcal{R}^D$ we can obtain 
\begin{equation}
k(x,y) = \langle k(\cdot,x), k(\cdot,y) \rangle \approx z(x)^Tz(y)
\end{equation}
where $x,y \in \mathcal{R}^d$. More specifically, \cite{ali2007} demonstrate the construction of feature spaces that uniformly approximate shift-invariant kernels with $D = O(d \epsilon^{-2} \log \frac{1}{\epsilon^2})$ where $\epsilon$ is the accuracy of approximation. However, as we will see later certain moment conditions need to be satisfied.\\

Bochner's Theorem provides the key observation behind such approximation. This classical theorem (Theorem 6.6 in \cite{Wendland}) is useful in several contexts where one deals with translation-invariant kernels $k$, i.e. $k(x,y) = \kappa(x-y)$. As well as constructing large-scale approximation to kernel methods \cite{ali2007}, it can also be used to determine whether a kernel is characteristic i.e. if the Fourier transform of a kernel is supported everywhere then the kernel is characteristic \citep{Bharath}.
%Bochner's Theorem
\begin{theorem}
\label{th:Bochner}
\textbf{Bochner's Theorem} \citep{Wendland} A continuous transition-invariant kernel $k$ on $\mathcal{R}^d$ is positive definite if and only if $k(\delta)$ is the Fourier transform of a non-negative measure. 
\end{theorem}

%Given enough training data, kernel machines can approximate any function or decision boundary arbitrarily well. Despite such an attractive feature, given the time allowed, these methods that involve kernel matrices can become infeasible to compute as the size of training data increases.

For a properly scaled transition-invariant kernel $k$, the theorem guarantees that its Fourier transform $\Gamma(w)$ is a non-negative measure on $\mathcal{R}^d$. Without loss of generality, $\Gamma$ is a probability distribution. Since we would like to approximate real-valued kernel matrices, let us consider the approximation which uses only real-valued features. $\kappa(x-y)$ can be written as : 
\begin{align}
\kappa(x-y) & = \int_{\mathcal{R}^d} \exp(iw^T(x-y)) d \Gamma(w) \\
          & = \int_{\mathcal{R}^d} \cos(w^T(x-y)) + i \sin(w^T(x-y)) d \Gamma(w) \\\label{eq:cos}  
          & = \int_{\mathcal{R}^d} \cos(w^T(x-y)) d \Gamma(w) \\     \label{eq:doubleangel}  
          & = \int_{\mathcal{R}^d} \left \{ \cos (w^Tx) \cos (w^Ty) + \sin(w^Tx)\sin(w^Ty) \right\} d \Gamma(w) 
\end{align}
provided that 
\begin{equation}
\label{eq: momentcon}
\mathbb{E}_\Gamma(w^Tw) < \infty.
\end{equation}
Note, \eqref{eq:cos} follows because kernels are real valued and \eqref{eq:doubleangel} uses the double angle formula for cosine. The random features can be computed by first sampling $\{w_j\}^D_{j=1} \iid \Gamma$ and then for $x_j \in \mathcal{R}^d$ with $j \in \{1, ..., n \}$, setting $z(x_j) = \sqrt{\frac{2}{D}}( \cos(w_1^Tx_j),\sin(w_1^Tx_j), ... , \cos(w_{\frac{D}{2}}^Tx_j), \sin(w_{\frac{D}{2}}^Tx_j))$ for $j \in \{1,...,n \}$. \\

Here, we deal with explicit feature space, and apply linear methods to approximate the Gram matrix through the covariance matrix $Z(x)^TZ(x)$ of dimension $D \times D$ where $Z(x)$ is the matrix of random features. Essentially, \eqref{eq: momentcon} guarantees that the second moment of the Fourier transform of this translational invariant kernel $k$ to be finite and hence ensure the uniform convergence of $z(x)^Tz(y)$ to $\kappa(x-y)$\cite{ali2007}.

%As a side note, the main focus of \cite{ali2007} is on another approach, but we will refer to the original paper \cite{ali2007} as it is not as relevant to our discussion here. 
% \textcolor{blue}{Each data point is therefore represented by random Fourier features. (this needs to be placed at an appropriate poitn..)}
 
 \subsubsection{RFF HSIC Estimator}
The derivation of the biased RFF HSIC estimator follows in the same manner as Section \ref{Sec:Nystats}. However, with the RFF approximations of the kernel matrices, \eqref{eq:NyHSIC} becomes: 
\begin{align} 
 & \hat \Xi_{RFF, \tilde k_\mathcal{X},\tilde k_\mathcal{Y}}(\mathbf{z}) \nonumber \\
   & = \left \Vert \frac{1}{m} \sum^m_{i=1} Z_x(x_i) Z_y(y_i)^T - \left ( \frac{1}{m} \sum^m_{i=1} Z_x(x_i) \right ) \left (\frac{1}{m} \sum^m_{i=1} Z_y(y_i) \right)^T \right \Vert^2_F \\
   &= \left \Vert \frac{1}{m}Z_x^T H  Z_y \right \Vert^2_F
 \end{align}
 where $Z_x \in \mathbb{R}^{m \times D_x}$ and $Z_y \in \mathbb{R}^{m \times D_y}$. Hence, when RFF estimators are substituted, the cross covariance operator is simply a $D_x \times D_y$ matrix. In the same way as the Nystr\"om HSIC estimator, the HSIC statistic computed using RFF is a HSIC statistic computed using a different kernel, i.e. one that is induced by the random features. It is worth noting that the analysis of convergence of such estimator can possibly be done similarly to the analysis by \cite{ErrorRFF} for MMD. However, we will leave this for future work. \\

 To use the RFF HSIC statistic in independence testing, the permutation approach and spectral approach in the previous section can be adopted for null distribution estimation with $\hat \Xi_{Ny, \tilde k_\mathcal{X},\tilde k_\mathcal{Y}}(\mathbf{z})$ replaced by $\hat \Xi_{RFF, \tilde k_\mathcal{X},\tilde k_\mathcal{Y}}(\mathbf{z})$. Just as the case with inducing points, the $\{w_j\}_{j=1}^{D_.}$ should be sampled each time independently for $X$ and $Y$ when the RFF approximations $Z_x$ and $Z_y$ needed to be computed. As a remark, the number of inducing points and the number of $w_j$s plays a similar role in both methods which controls the trade off between computational complexity and statistical power. In practice, as we will demonstrate in the next section, such number can be much smaller than the size of the dataset without compromising the performance. 
 %When we estimate the eigenvalues with RFF, the HSIC statistic that is subsequently computed is essentially a HSIC statistic computed using the kernel induced by the random features. Note, this kernel will be different from the original kernel we used to compute the Gram matrix before projecting the data through the randomised feature map. Just in the same way as we previously discussed, the linear time estimator follows that of the most general case as Expression.\ref{ex: DisBtest}. The null distributions of the quadratic time statistics take the form of Expression.\ref{eq:HSICasy} and \ref{eq:HISCdis} for the biased and unbiased statistics respectively. However, now, as the RKHS is finite dimensional, both become a finite sum of chi-square random variables. 
 
\section{Experiments}
\label{sec:experiments}
In this section, we present three synthetic data experiments to study the behaviour of our large scale HSIC tests.  The main experiment is on a challenging non-linear low signal-to-noise ratio dependence dataset to assess the numerical performance amongst the large scale HSIC tests. To investigate the performance of these test in a small scale, we further conduct linear and sine dependence experiments to compare with currently established methods for independence testing. 
Throughout this section, we set the significance level of the hypothesis test to be $\alpha = 0.05$. Both Type I and Type II errors are calculated based on 100 trials. The $95\%$ confidence intervals are computed based on normality assumption, i.e. $\hat \mu \pm 1.96 \sqrt{\frac{\hat \mu (1- \hat \mu)}{100}}$, where $\hat \mu$ is the estimate for the statistical power.

\subsection{Simple Linear Experiment}

We begin with an investigation of the performance of our methods on a toy example with a small number of observations, in order to check the agreements between large scale approximation methods we proposed and the exact methods where they are still feasible. Towards this aim, we consider a simple linear dependence experiment, but where the dependence between the response $Y$ and the input $X$ is in only a single dimension of $X$. In particular, 
$$X \sim \mathcal{N}(0,I_d)\ \ \  \text{and}\ \ \  Y= X_1 + Z$$ 
where $d$ is the dimensionality of data vector $X$ and $X_1$ indicate the first dimension of $X$. The noise $Z$ is independent standard Gaussian noise.  We would like to compare methods based on HSIC to a method based on Pearson's Correlation which is explicitly aimed at linear dependence and should give strongest performance. However, as the latter cannot be directly applied to multivariate data, we consider a SubCorr statistic: SubCorr $=\frac{1}{d}\sum^d_{i=1} \text{Corr}(Y,X_i)^2$ where Corr($Y,X_i$) is the Pearson's correlation between $Y$ and the $i^{th}$ dimension of $X$ . In addition, we will also consider SubHSIC statistic: SubHSIC$=\frac{1}{d}\sum^d_{i=1} \text{HSIC}(Y,X_i)^2$. For these two methods, we will use a permutation approach as their distributions are not immediately clear.

\begin{figure}[h] 
\begin{center}
\leftskip=-1.55cm
\includegraphics[width=1.27\textwidth]{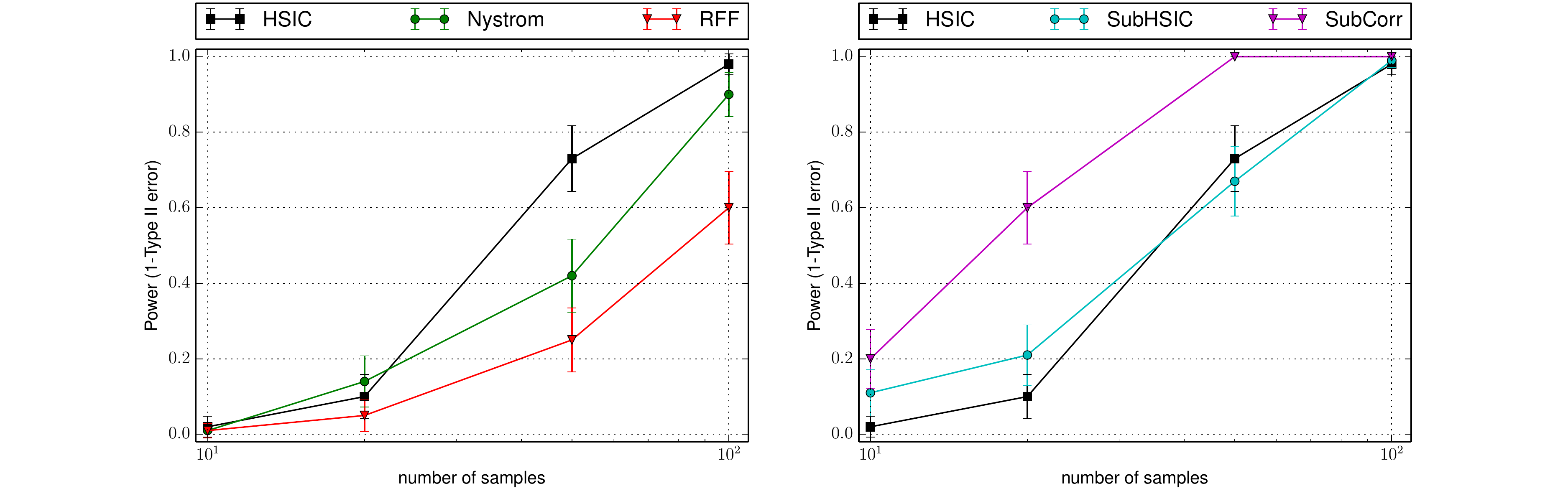}
\caption{Simple linear experiment for $d=10$. Left: comparing HSIC spectral approach with Nystr\"om spectral method and RFF spectral method. Right: 
HSIC spectral approach with SubHSIC and SubCorr.}
\label{fig: Linear}
\end{center}
\end{figure}

In Fig \ref{fig: Linear}, the dimension of $X$ is set to be 10. Both the number of random features in RFF and the number of inducing variables in Nystr\"om are set to 10. We do not use the block-based method as the sample sizes are small. From Fig \ref{fig: Linear} (right), we see that SubCorr yields the highest power as expected. HSIC and SubHSIC with Gaussian median-heuristic kernels perform similarly though, with all three giving the power of 1 at the sample size of 100. On the other hand, Fig \ref{fig: Linear} (left) shows that the two large scale methods are still able to detect the dependence at these small sample sizes, even though there is some loss in power in comparison to HSIC and they would require a larger sample size. As we will see, this requirement for a larger sample size will be offset by a much lower computational cost in large-scale examples.  %This experiment gives us a good indication of how much we actually lose by using these proposed large scale HSIC tests. In fact, we do not lose much. 
 %The two large scale methods give the worse performance at almost all numbers of samples with Nystr\"om method performing slightly better than the RFF method. 

\subsection{Sine Dependence Experiment }
We now consider a more challenging nonlinear dependence experiment to investigate time vs power tradeoffs of large-scale tests. %Further, to investigate the performance of the large scale methods in comparison to the distance correlation (dCor) \cite{Cordis2007,SR2009}, we consider a sine dependency example, which is slightly less difficult comparing to the sign example above.
The dataset consists of a sample of size $m$ generated i.i.d. according to: 
$$X \sim \mathcal{N}(0,I_d)\ \ \  \text{and}\ \ \  Y= 20\sin(4\pi(X_1^2+X_2^2)) + Z$$ 
where $d$ is the dimensionality of data vector $X$, $X_i$ indicates the $i^{\text{th}}$ dimension of $X$ and $Z\sim\mathcal N(0,1)$. In addition to HSIC and its large scale versions, we will also consider dCor \cite{Cordis2007,SR2009} - which can be formulated in terms of HSIC using Brownian Kernel with parameter $H=0.5$ \cite[Appendix A]{SejSriGreFuku13}. In addition, we will consider dCor using the Gaussian kernel with median heuristic bandwidth parameter.\\ %which we introduce  the shorthand notation GdCor.  \\

\begin{figure}[h] 
\begin{center}
\includegraphics[width=0.85\textwidth]{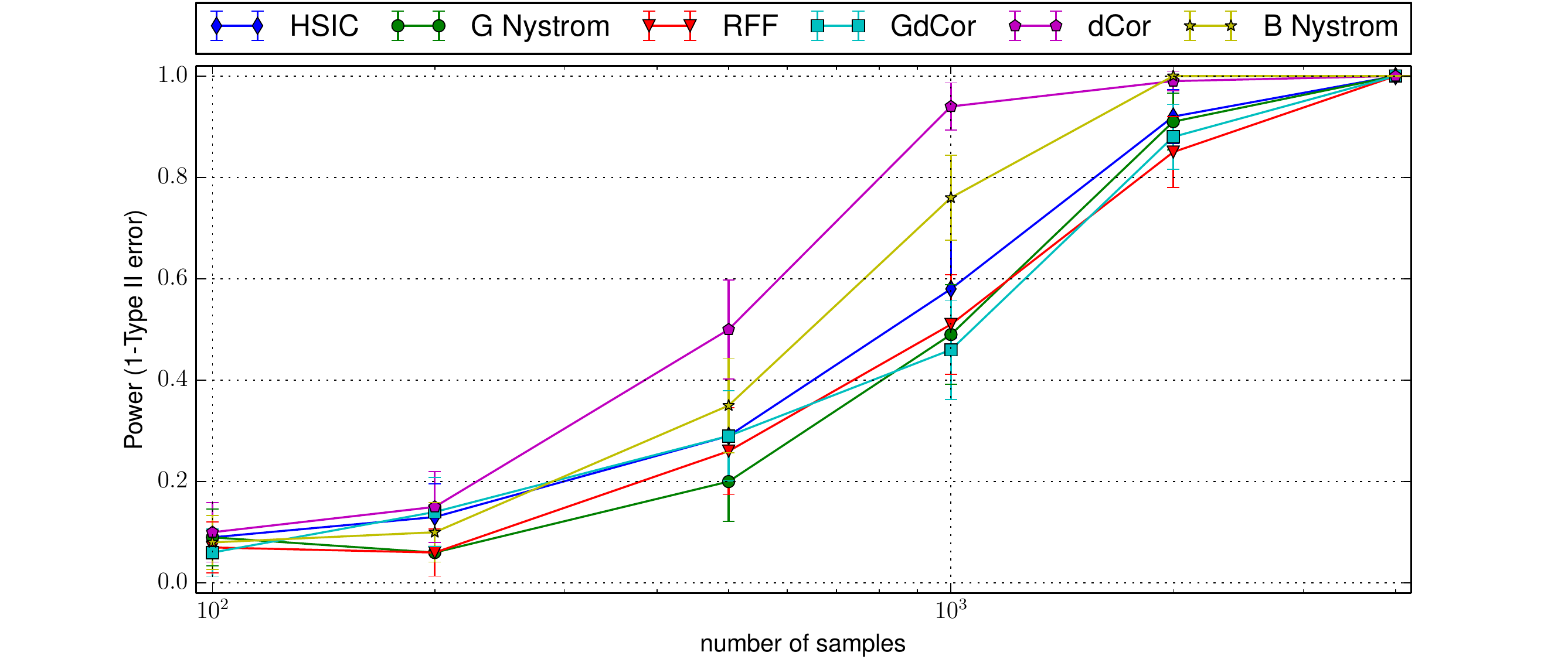}
\caption{Sine dependence experiment for $d=2$ comparing HSIC spectral approach, Nystr\"om spectral method (G: Gaussian RBF kernel with median heuristic; B: Brownian kernel with $H=0.5$ ), RFF  spectral method, dCor and GdCor (dCor with Gaussian RBF kernel median heuristic).}
\label{fig: Sine}
\end{center}
\end{figure}

\begin{figure}[h] 
\begin{center}
\includegraphics[width=0.85\textwidth]{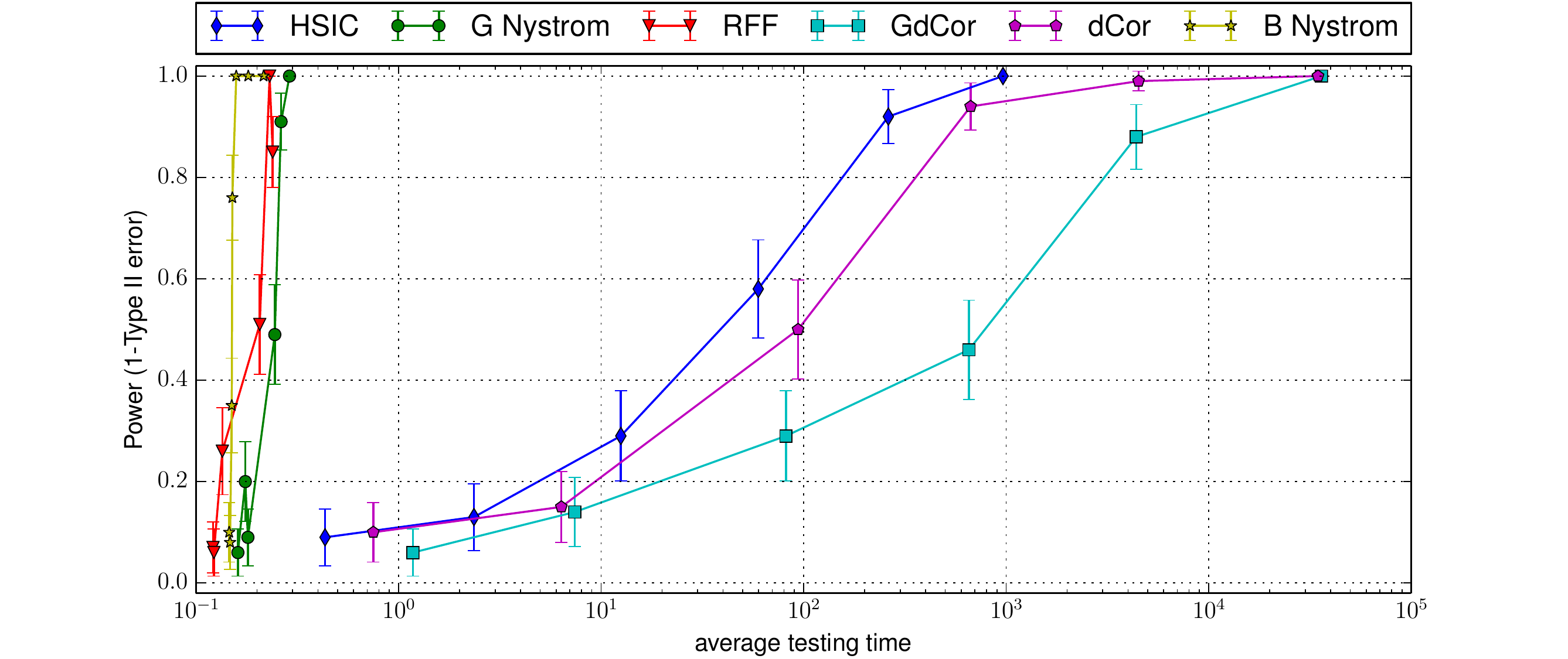}
\caption{The corresponding average testing time plot for the sine dependence experiment for $d=2$.}
\label{fig: SineT}
\end{center}
\end{figure}

The number of random Fourier features and the number of inducing variables are both set to 50. For RFF, we use the Gaussian kernel with median heuristic bandwidth parameter, while for Nystr\"om we in addition use the Brownian kernel with $H=0.5$ (note that RFF is not applicable to this kernel as it is not translation-invariant). At these still relatively small sample sizes, block-based approach gave poor performance. From Fig. \ref{fig: Sine}, dCor clearly outperforms the other methods with the Brownian kernel Nystr\"om method giving the closest performance in terms of power. Reassuringly, the four methods using Gaussian kernel all give very similar power performance. Fig. \ref{fig: SineT}, however, tells a very different story - the large-scale methods all reach the power of 1 in a test time which is several orders of magnitude smaller, demonstrating the utility of the introduced tests.

%In simple problems like that of a linear dependence, the large scale methods experience a lost in power comparing to the exact methods. However as the problem gets progressively more difficult (e.g. sine dependence and the low signal-to-noise dependence), the large scale approximations provide similar performance whilst significantly reduced the computational complexity in memory and time.  

\subsection{Large Scale Experiment}

We would now like to more closely compare the performance of the proposed large scale HSIC tests with each other -- at sample sizes where standard HSIC / dCor approaches are no longer feasible. We consider a challenging non-linear and low signal-to-noise ratio experiment, where a sample of size $m$ is generated i.i.d. according to: 
$$X \sim \mathcal{N}(0,I_d)\ \ \  \text{and}\ \ \  Y = \sqrt{\frac{2}{d}}\sum^{d/2}_{j=1}\text{sign}(X_{2j-1}X_{2j})|Z_j| + Z_{\frac{d}{2}+1} $$
where $d$ is the dimensionality of the data set $X$ and $Z \sim \mathcal{N}(0,I_{\frac{d}{2}+1})$. Note that $Y$ is independent of each individual dimension of $X$ and that the dependence is non-linear. For $d = 50$ and $100$, we would like to explore the relationship between the test power across a different number of samples $m = \{ 10^5, 2\times10^5, 5\times 10^5, 10^6, 2\times 10^6,5\times 10^6, 10^7 \}$.  The number of random features, inducing variables and block size are all set to 200 so that their computational cost is comparable. Gaussian RBF kernel with median heuristic is used in all cases. For RFF and Nystr\"om methods, we used the spectral approach to estimate the null distribution. \\

\begin{figure}[h] 
\begin{center}
\includegraphics[width=\textwidth]{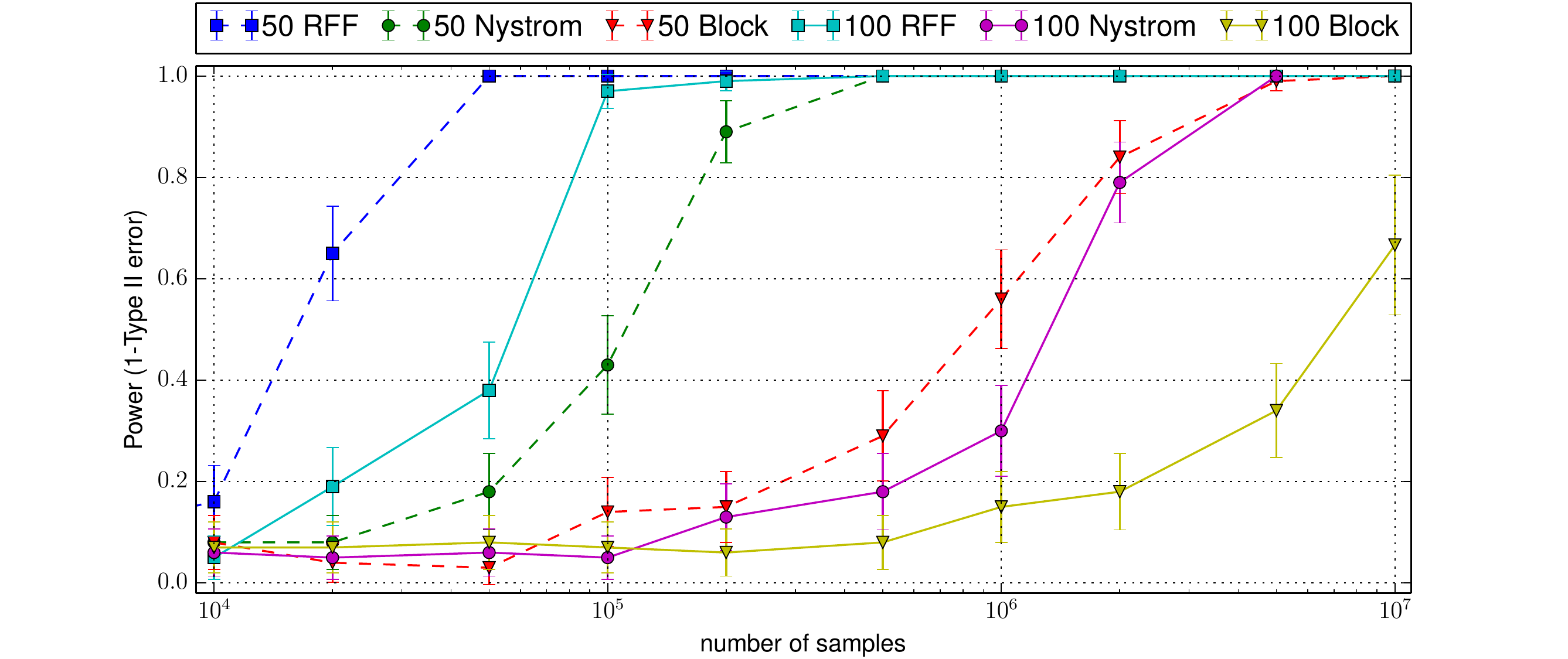}
\caption{Large Scale Experiment: The statistical power comparison between the three large scale independence testing methods based on 100 trials. Dotted line: $d = 50$; solid line: $d=100$. The $95\%$ confidence intervals are computed based on normality assumption, i.e. $\hat \mu \pm 1.96 \sqrt{\frac{\hat \mu (1- \hat \mu)}{100}}$, where $\hat \mu$ is the estimate for the statistical power.}
\label{fig: LSPower}
\end{center}
\end{figure}

\begin{figure}[h] 
\begin{center}
\includegraphics[width=\textwidth]{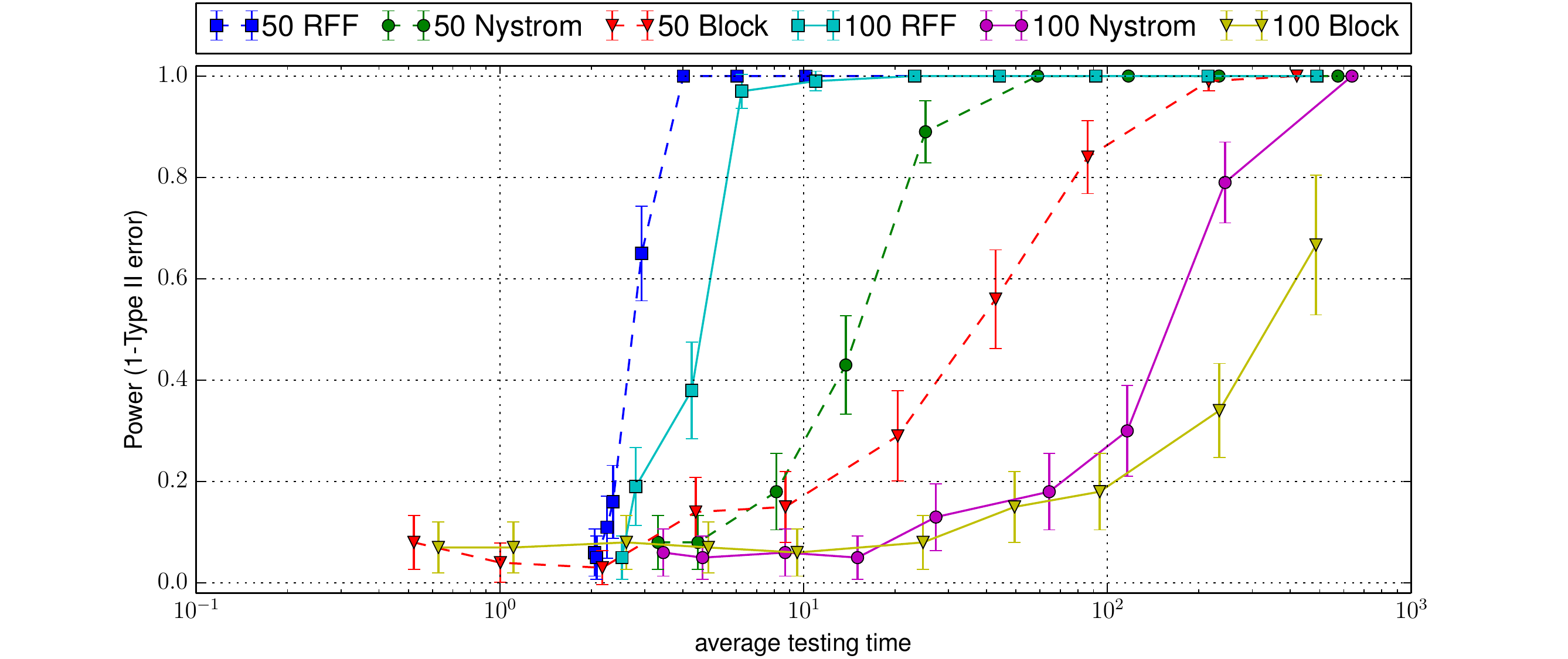} %LSTime2
\caption{Large Scale Experiment: The average testing time comparison between the three large scale independence testing methods. Dotted line: $d = 50$; solid line: $d=100$.}
\label{fig: LSTime}
\end{center}
\end{figure}

Fig. \ref{fig: LSPower} is a plot of the test power against the number of samples whereas Fig. \ref{fig: LSTime} is a plot of the test power against average testing time.  It is clear that for both $d=50$ and $ d = 100$, the RFF method gives the best performance in power for a fixed number of samples, followed by the Nystr\"om method and then by the block-based approach. The RFF method is able to achieve zero type II error (i.e. no failure to reject a false null) with 5$\times 10^4$ samples for $d=50$ and 5$\times 10^5$ samples for $d=100$, while the Nystr\"om method has a $80\%$ false negative rate at these sample sizes. The power vs time plot in \ref{fig: LSTime} gives a similar picture as Fig. \ref{fig: LSPower} confirming the superiority of the RFF method on this example.   \\

\section{Discussion and Conclusions}

We have proposed three novel large scale estimators of HSIC, a kernel-based nonparametric dependence measure -- these are the block-based estimator, the Nystr\"om estimator and the RFF estimator. We subsequently established suitable independence testing procedures for each method -- by taking advantage of the normal asymptotic null distribution of the block-based estimator and by employing an approach that directly estimates the eigenvalues appearing in the asymptotic null distribution for the Nystr\"om and RFF methods. All three tests significantly reduce computational complexity in memory and time over the standard HSIC-based test. We verified the validity of our large-scale testing methods and its favourable tradeoffs between testing power and computational complexity on challenging high-dimensional synthetic data. We have observed that RFF and Nystr\"om approaches have considerable advantages over the block-based test. Several further extensions can be studied: the developed large-scale approximations are readily applicable to three-variable interaction testing \cite{SejGreBer2013x}, conditional independence testing \cite{Fukumizu08kernelmeasures} as well as application in causal discovery \cite{ZhPeJanSch11, Flaxman2015}. Moreover, the RFF HSIC approach can be extended using the additional smoothing of characteristic function representations similarly to the approach of \cite{ChwRamSejGre2015} in the context of two-sample testing.

\bibliographystyle{spmpsci}      % mathematics and physical sciences
\bibliography{bibLS}

\appendix
\section{Proof of Theorem \ref{th:asynull}} \label{sec:Appendix}

We note that \cite{L2013} gives a proof to a similar theorem (Theorem 2.7 in \cite{L2013}) regarding generalised versions of distance covariance (dCov). We closely follow the steps given in Proposition 2.6 and Theorem 2.7 of \cite{L2013}. However, the proof provided here is tailored to the kernel view of HSIC/dCov duality \cite{SejSriGreFuku13} and it is slightly more general in that it applies to any semimetric (rather than metric) of negative type. Unless stated otherwise, the notation follows that used in the main part of this paper.  \\

%For notational convenience, we denote $P_X$ by $\mu$, $P_Y$ by $\nu$ and $P_{XY}$ by $\theta$. 
Recall from the main part of this paper that the existence of the HSIC statistics $\Xi_{k_\mathcal{X},k_\mathcal{Y}}(\mathbf{z})$ defined in \eqref{eq:HSIC_as_expectations} requires the marginal distributions to have finite first moment with respect to the kernels, i.e. $P_X \in \mathcal{M}^1_{k_{\mathcal X}}(\mathcal{X})$ and $P_Y \in \mathcal{M}^1_{k_{\mathcal Y}}(\mathcal{Y})$. By Proposition 20 of \cite{SejSriGreFuku13}, this translates directly into finite first moment conditions with respect to the semimetrics:  $P_X \in \mathcal{M}^1_{d_x}(\mathcal{X})$ and $P_Y \in \mathcal{M}^1_{d_y}(\mathcal{Y})$ when $k_{\mathcal X}$ generates $d_x$ and $k_{\mathcal Y}$ generates $d_y$. \\

%Let $(\mathcal{X}, d_\mu)$ and $(\mathcal{Y},d_\nu)$ be semimetric spaces of negative type.

More specifically, a valid semimetric $d$ of negative type on $\mathcal{Z}$ generated by a non-degenerate kernel $k$ on $\mathcal{Z}$ can be written as $d(z,z') = k(z,z) + k(z',z')-2k(z,z')$ (Corollary 16 \cite{SejSriGreFuku13}).   Then, such semimetric $d_x$ centred at the the probability measure $P_X$ defined on $\mathcal{X}$ is  %given by \cite{L2013} as 
\begin{align}
\label{eq: du}
d_{P_X} (x,x') :&= d_x(x,x') - \int d_x(x,x') dP_X(x')  \nonumber\\
&- \int d_x(x',x) dP_X(x) + \int d_x(x,x') dP_X^2(x,x') 
\end{align}
Similarly, $d_{P_Y} (y, y')$ is the semimetric centred at the probability measure $P_Y$ defined on $\mathcal{Y}$. If we substitute the kernel representation of the semimetric for $X$ and $Y$ respectively into $d_{P_X} (x,x')$ and $d_{P_Y} (y,y')$, we obtain the following: 
\begin{equation}
d_{P_X}(x, x') = -2 \tilde k_{P_X}(x,x') \ and \ d_{P_Y}(y, y') = -2 \tilde k_{P_Y}(y,y') 
\end{equation} 
where $\tilde k_{P_X}(x,x')$ and $ \tilde k_{P_Y}(y,y') $ are defined in \eqref{eq: centred kernel}. \\ %Note that $\mathbb{E}(d_\mu(X, X')d_\nu(Y, Y'))$ is well defined under the imposed moment condition \cite{SejSriGreFuku13}. \\

Let $(X^i,Y^i) \sim \theta$ be independent for $ i \in \{1,2,...,6\}$, we introduce the ``core" defined in \cite{L2013}, 
\begin{equation}
\label{eq: h}
h((X^1,Y^1),(X^2,Y^2), ... , (X^6,Y^6)) := f(X^1,X^2,X^3,X^4)f(Y^1,Y^2,Y^5,Y^6) 
\end{equation}
where for $z_i \in \mathcal{X}$ or $z_i \in \mathcal{Y}$:
\begin{align}
f(z_1, z_2, z_3, z_4) :&= d.(z_1,z_2) - d.(z_1,z_3) - d.(z_2, z_4) + d.(z_3,z_4) \\
& = -2 [k.(z_1,z_2) - k.(z_1,z_3) - k.(z_2, z_4) + k.(z_3,z_4)] \label{eq: dtok}
\end{align}
The second line follows from the relationship between $d$ and $k$ where $k_{P_X}$ is used for $z_i \in \mathcal{X}$ and $k_{P_Y}$ for $z_i \in \mathcal{Y}$.\\

%\textcolor{blue}{???? I am not sure if this following paragraph is absolutely necessary... Since $\mathbb{E}(d_\mu(X, X')d_\nu(Y, Y'))$ is well defined under the imposed moment condition \cite{SejSriGreFuku13}, then 
%\begin{align}
%\mathbb{E}(h((X^1,Y^1),(X^2,Y^2), ... , (X^6,Y^6))) 
%&=  \mathbb{E}(d_\mu(X, X')d_\nu(Y, Y')) \\
%&= 4 \mathbb{E}(\tilde k_\mu(X,X') \tilde k_\nu(Y,Y')) \\
%&= 4HISC_b (\mathbf{Z}; k_\mu,k_\nu) \label{eq:hHSIC} 
%\end{align}
%Actually, based on Dino's paper, we have that the asymptotic distribution of HSICb only require the finite first moment of marginals. Our proof here should directly prove in that context! }

In fact, we can prove that the expectation of \eqref{eq: h} is four times the HSIC of $X$ and $Y$. To see this, we first need to show that such expectation is well defined. Indeed, note that for a valid semimetric $d$ of negative type on $\mathcal{Z}$, $\sqrt{d(z,z')} = ||k(\cdot,z) - k(.,z') ||_{\mathcal{H}_k}$, then the following inequality holds:
\begin{equation}
d(x,y) \leq d(x,z) + d(y,z) + 2\sqrt{d(x,z)d(y,z)} \ \  \forall x,y,z \in \mathcal{Z}
\end{equation}
It then follows that 
\begin{align*}
&|f(z_1,z_2,z_3,z_4)| \\
\leq & \ 2d(z_2,z_3) + 2 \sqrt{d(z_1,z_3)d(z_2,z_3)} +2 \sqrt{d(z_2,z_3)d(z_2,z_4)} \\
\leq & \ 2d(z_2,z_3) + 2 \max \{d(z_1,z_3),d(z_2,z_3)\} +2 \max \{d(z_2,z_3),d(z_2,z_4)\} \\
 \leq & \ 4 \left [ k(z_2,z_3) +  \max \{k(z_1,z_3),k(z_2,z_3)\} + \max \{k(z_2,z_3),k(z_2,z_4)\} \right ] \\
:= & \ g_1(z_1,z_2,z_3,z_4) 
\end{align*}
and that 
\begin{align*}
&|f(z_1,z_2,z_3,z_4)| \\
\leq & \ 2d(z_1,z_4) + 2 \sqrt{d(z_2,z_4)d(z_1,z_4)} +2 \sqrt{d(z_1,z_3)d(z_1,z_4)} \\
\leq & \ 2d(z_1,z_4) + 2 \max\{d(z_2,z_4),d(z_1,z_4)\} +2 \max\{d(z_1,z_3),d(z_1,z_4)\} \\
\leq & \ 4 \left [ k(z_1,z_4) + 2 \max\{k(z_2,z_4),k(z_1,z_4)\} +2 \max\{k(z_1,z_3),k(z_1,z_4)\}  \right ] \\
:= & \ g_2(z_1,z_2,z_3,z_4)
\end{align*}
Hence, replacing the $z_i$ in $g_1$ with $X^i$ and the $z_i$ in $g_2$ with $Y^i$, 
\begin{equation*}
|h((X^1,Y^1),(X^2,Y^2), ... , (X^6,Y^6))| \leq g_1(X^1,X^2,X^3,X^4) g_2(Y^1,Y^2,Y^5,Y^6).
\end{equation*}
%$g_1$ and $g_2$ are both integrable and they are independent under the null hypothesis. Hence, $h$ is integrable. 
{Since the marginal distributions have finite first moments with respect to the kernels, then each of the terms in $g_1$ and $g_2$ is integrable and hence $g_1$ and $g_2$ are integrable. Moreover, since the marginal distributions have finite second moments with respect to the kernels, then the joint distribution satisfies $P_{XY} \in \mathcal{M}^1_{k_{\mathcal X} \otimes k_{\mathcal Y}}(\mathcal{X}\times \mathcal{Y})$}. Therefore, $h$ is integrable. Subsequently, by taking the expectation and utilising Fubini's theorem, we obtain that 
\begin{align}
\mathbb{E}(h((X^1,Y^1),(X^2,Y^2), ... , (X^6,Y^6))) 
&= 4 \mathbb{E}(\tilde k_{P_X}(X,X') \tilde k_{P_Y}(Y,Y')) \\
&= 4 \Xi_{k_\mathcal{X},k_\mathcal{Y}}(\mathbf{z}) \label{eq:hHSIC} 
\end{align}
which is 4 times the HSIC. \\

%Up until this point, we have not use the fact that we are under the null hypothesis. 
%\textcolor{red}{$HISC_b (\mathbf{z}; k_\mu,k_\nu)$ Note that $\mathbf{Z}$ is the random variable $((X,Y),(X',Y'))$.\\}

In order to use the theory of degenerate V-statistics to obtain the asymptotic distribution, we need to consider the symmetrised version of $h$, which we define as follows 
$$ \bar h ((X^1,Y^1), ... , (X^6,Y^6)) := \frac{1}{6!} \sum_{\sigma \in A} h((X^{\sigma(1)},Y^{\sigma(1)}), ... , (X^{\sigma(6)}, Y^{\sigma(6)})) $$
where $A$ denotes the set of all permutations of $\{1,...,6\}$. Then, under the null hypothesis of independence $P_{XY} = P_X \times P_Y $,
\begin{align*}
\bar h_2((x,y),(x',y')) :&=\mathbb{E} [\bar h ((x,y),(x',y'),(X^3,Y^3),... , (X^6,Y^6))] \\
&= \frac{4}{15} \tilde k_\mu(x,x') \tilde k_\nu(y,y')
\end{align*} 
If we fix the first two positions to be $\{1,2\}$ and randomly permute the rest, we obtain 24 different combinations. Similarly if we fix the first two positions to be $\{2,1\}$, we also obtain 24 different combinations. Some algebraic manipulation shows that these are the combinations that gives the expectation of $h$ to be $\tilde k_\mu (x,x') \tilde k_\nu (y,y').$ In fact, these are the only combinations as when either $\{1\}$ or $\{2\}$ or both are not in the first two positions, the expectation of $h$ is zero and all terms cancel out.  \\

Another important condition to check is that $\bar h_2$ has finite second moment.  It was shown earlier that  $h((X^1,Y^1), ... , (X^6,Y^6))$ of \eqref{eq: h} is integrable. Subsequently, $f(X^1,X^2,X^3,X^4)$ has finite second moment. Hence, $h((X^1,Y^1), ... , (X^6,Y^6))$ has finite second moment under the null hypothesis. Additionally, by Jensen's inequality, $\mathbb{E}(|\bar h_2 ((X,Y),(X,Y))|) \leq \{\mathbb{E}((\bar h_2 ((X,Y),(X,Y)))^2) \}^{1/2} < \infty$. \\

Hence, by Theorem B in Chapter 6 of \cite{serfling}, which says 
$$
m \left (\frac{1}{m^2} \sum_{i,j}^m \bar h_2((x_i,y_i),(x_j,y_j)) \right ) \xrightarrow{D} \sum^\infty_{r=1} \gamma_r Z^2_r 
$$
as the sample size $m \rightarrow \infty$, we obtain that 
\begin{equation}
m \left (\frac{1}{m^2} \sum_{i,j}^m \tilde k_{P_X}(x_i,x_j) \tilde k_{P_Y} (y_i,y_j) \right )  \xrightarrow{D} \sum^\infty_{r=1} \gamma_r Z^2_r 
\end{equation}
with $Z_r \iid \mathcal{N}(0,1) \  \forall r$ and $\{ \gamma_r \}^\infty_{r=1}$ are the eigenvalues of the operator $S_{\tilde k }$: $L^2_\theta (\mathcal{X} \times \mathcal{Y}) \rightarrow L^2_\theta (\mathcal{X} \times \mathcal{Y})$ defined as: 
$$
S_{\tilde k } g(x,y) = \int_{\mathcal{X} \times \mathcal{Y}}  \tilde k_{P_X} (x,x') \tilde k_{P_Y} (y,y') g(x',y') d\theta(x',y')
$$
Note, since under the null hypothesis $P_{XY} = P_X \times P_Y$, the above operator is given by the tensor product of $S_{\tilde k_{P_X}}$ and $S_{\tilde k_{P_Y}}$ (Remark 2.9 \cite{L2013}). Therefore $\{ \gamma_r \}^\infty_{r=1}$ are the products of the eigenvalues of these two operators. By noting that $\frac{1}{m^2} \sum_{i,j} \tilde k_{P_X}(x_i,x_j) \tilde k_{P_Y} (y_i,y_j)$ is exactly $\Xi_{b,k_\mathcal{X},k_\mathcal{Y}}(\mathbf{Z})$, i.e. the V-statistics with the kernel $\bar h_2((x,y),(x',y'))$ We obtained the desired asymptotic distribution.

\end{document}